\documentstyle[11pt,epsfig]{article}

\renewenvironment{thebibliography}[1]
{\normalsize
 \begin{list}{[\arabic{enumi}]}
{\usecounter{enumi} \setlength{\parsep}{0pt}
 \setlength{\itemsep}{3pt} \settowidth{\labelwidth}{[#1]}
 \sloppy}}
{\end{list}}

\newcommand{\pr}{\hspace{\parindent}}

\begin{document}

\hfill\vbox{\baselineskip14pt
            \hbox{\bf YUMS 97-7}
            \hbox{\bf SNUTP 97-035}
            \hbox{\bf \today}}

\baselineskip20pt

\begin{center}
{\Large{\bf $CP$ Violation in the Top-Quark Pair Production\\
            at a Next Linear Collider}}         
\end{center} 

\vskip 1cm
\begin{center}
\large M.S.~Baek, S.Y.~Choi and C.S.~Kim 
\end{center}

\begin{center}
Department of Physics, Yonsei University, Seoul 120-749, Korea
\end{center}

\vskip 2cm
\begin{center} 
\large Abstract
\end{center}

\baselineskip=16pt

We provide a detailed, model-independent, study for  $CP$ violation effects 
due to the $T$-odd top-quark electric dipole moment (EDM) and weak dipole 
moment (WDM) in the top-quark pair production via $e^+e^-$ and two-photon 
annihilation at a next $e^+e^-$ linear collider (NLC).
There are two methods in detecting $CP$ violation effects in these processes. 
One method makes use of measurements of various spin correlations in the final 
decay products of the produced top-quark pair, while the other
is to measure various $CP$-odd polarization asymmetry effects of the initial 
states. In the $e^+e^-$ case only the first method can be used, and 
in the $\gamma\gamma$ case both methods can be employed. We provide
a complete classification of angular correlations of the $t$ and $\bar{t}$
decay products under $CP$ and $CP\tilde{T}$ which greatly faciliate $CP$
tests in the $e^+e^-$ mode. Concentrating on the second method with 
the Compton back-scattered high-energetic laser light off the electron 
or positron beam in the two-photon mode, we construct two $CP$-odd 
and $CP\tilde{T}$-even initial polarization configurations and apply them 
to investigating $CP$-violating effects due to the top-quark EDM.
With a typical set of experimental parameters at the NLC, we compare 
the 1-$\sigma$ sensitivities to the top-quark EDM and WDM in the $e^+e^-$
mode and the two-photon mode.
Some model expectation values of the $T$-odd parameters are 
compared with the results.

\baselineskip=20pt
\normalsize

\newpage

\section{Introduction}
\label{sec:Top_intro}

\pr
Precision measurements of various production and decay modes of the top
quark are expected to provide useful information on physics beyond the
SM. Testing new physics in observables which are
sensitive to $CP$ violation seems especially promising. As the top quark
hardly mixes with other generations, the GIM mechanism of unitarity
constraints leads to negligibly small effects of $CP$ violation in the
SM. Thus, observation of $CP$ non-invariance in top-quark physics would
definitely be a signal for physics beyond the SM.

An important property of heavy top ($m_t\sim 175$ GeV)\cite{PDG96} 
is that
hadronization of the top quark can be neglected to a good approximation
because on average it decays before it can form hadronic
bound states\cite{BDKKZ}. This implies in particular that spin effects, 
for instance spin correlations between the produced $t$ and 
$\bar{t}$ quarks are not be severely distorted by hadronization. 
These spin effects can be analyzed through the distributions and angular
correlations of the weak decay products of the $t$ and $\bar{t}$
quarks. Moreover, these effects can be calculated in perturbation theory.
Hence they provide an additional means of testing the SM predictions and 
of searching for possible new physics effects in top quark production
and decay.

The $\gamma t\bar{t}$ coupling consists of the SM tree-level and 
the magnetic dipole moment (MDM) couplings as well as the EDM coupling. 
Likewise, in addition to the tree-level SM $Zt\bar{t}$ coupling, we have 
the analogous $Z$ MDM and $Z$ EDM couplings, the latter of which
is called the top-quark WDM. In both cases these couplings may
have imaginary parts. 
The MDM-like couplings are present in the SM at the one-loop level.
On the other hand, the EDM-like couplings violate $CP$ and,
due to the structure of the SM, they are only present perturbatively
in the SM at the three loop level. In some extensions to the SM, 
however, EDM couplings may be present at lower order in perturbation 
theory. Some models\cite{WbMs} which can give relatively large fermion 
EDM's include
left-right symmetric theories, additional Higgs multiplets,
supersymmetry, and composite models. Neglecting the MDM-like couplings, 
we consider both the $T$-odd top-quark EDM and WDM in the reactions
$e^+e^-\rightarrow t\bar{t}$ and $\gamma\gamma \rightarrow t\bar{t}$
at NLC\cite{NLC}. 

Previously $CP$ violation in the process $e^+e^-\rightarrow t\bar{t}$ 
has been extensively investigated. 
Those previous works can be classified in two categories according to
their own emphasized aspects: (i) the classification\cite{WbAb,BN,KLY,AA} 
of spin correlations of the $t$ and $\bar{t}$ decay products without 
electron beam polarization, and (ii) the use of a few typical $CP$-odd 
observables with electron beam polarization\cite{CR,PR}. 
In the first class they have constructed
all the $CP$-odd observables according to their ranks. However, since 
the $t$ and $\bar{t}$ are spin-1/2 particles,  the $CP$-odd 
spin correlation only up to rank-two can appear in the process.
Therefore, all the constructed $CP$-odd observables previously investigated
are not linearly independent. In the present work we completely define
all the linearly-independent $CP$-odd correlations by which all other $CP$-odd
correlations can be expressed.  In the second class, it has been shown 
that electron beam polarization is very crucial for a few specialized 
$CP$-odd correlations. We extend their works to investigate 
which $CP$-odd correlations depend crucially on electron polarization
and which ones do not. After expressing all the strongly-dependent
$CP$-odd correlations in terms of the linearly-independent correlations
we can provide simple explanations for why those specialized 
observables depend crucially on electron beam polarization.

Detailed studies have been performed mainly in the processes 
$e^+e^-\rightarrow t\bar{t}$ including general 
studies of $tt\gamma$, $ttZ$ and $tbW$ couplings\cite{WbAb,BN,KLY,AA}
previously. 
A photon linear collider (PLC), employing polarized photons by the Compton
back-scattering of polarized laser light on electron/positron beams of NLC, 
enables us to measure the $tt\gamma$ and 
$tbW$ couplings and to investigate the possibility of extracting
the effective couplings of the top quark to the photon.

We can employ two methods to extract the top-quark effective couplings 
at a PLC. One method makes use of the quasi-freely decaying 
property\cite{BDKKZ}
of the top quark by measuring various spin correlations in the
$t\bar{t}$ final system, $(bW^+)(\bar{b}W^-)$ or 
$(bf_1\bar{f}_2)(\bar{b}f_3\bar{f}_4)$.
The other method is to employ linearly-polarized photon beams 
to measure various polarization asymmetries of the initial 
states. It is, of course, possible to combine the two
methods. The former technique is essentially same as
that employed in $e^+e^-$ collisions\cite{BN,KLY,AA}
with one difference; in $e^+e^-$ collisions the spin of the 
$t\bar{t}$ system is restricted to $J=1$, while in photon
fusion $J=0$ or $J\geq 2$ is allowed.

Section~2 is devoted to the introduction of the top-quark EDM and WDM,
and to some model expectations for the $CP$-odd parameters.
In Section~3 we classify all angular dependences and angular correlations
of $t$ and $\bar{t}$ decay products in the $e^+e^-$ mode under $CP$ and 
$CP\tilde{T}$ transformations, where $\tilde{T}$ is the ``naive" time reversal
operation which flips particle momenta and spins but does not 
interchange initial and final states. Then we apply the $CP$-odd
angular correlations to probing $CP$ violation due to the top-quark
EDM and WDM in the $e^+e^-$ mode by
considering the polarized electron case as well as the unpolarized 
electron case. 

In Section~4 we give a detailed description of the energy spectrum
and polarization of the high-energy Compton backscattered light.
The study of $CP$ violation in the two-photon mode\cite{ScKh}
is extended in Section~5, where we construct two $CP$-odd and 
$CP\tilde{T}$-even initial photon polarization configurations, and
apply them to obtain the 1-$\sigma$ sensitivities of the top-quark EDM without 
the detailed information on the complicated $t$ and $\bar{t}$ decay patterns. 

After comparing the 1-$\sigma$ seinsitivities to the top-quark EDM 
in the $e^+e^-$ mode and the two-photon mode, we close Section~6 with 
some prospects for futher studies related with $CP$ violation.

The Appendices are devoted to the definition and explicit form
of the angular distributions ${\cal P}_{\alpha X}$ and the
definition of the angular correlations ${\cal D}_\alpha$ and 
${\cal D}^\prime_\beta$.

\section{Top-quark EDM}
\label{sec:top EDM}

\pr
One of the most commonly studied $CP$-violating operators 
is the EDM of a fermion and its 
generalizations to weak and strong couplings. The most general matrix 
element of the electromagnetic current between two top-quark spinors contain
$T$-odd term:
\begin{equation}
\langle t| j^{em}_\mu |t \rangle =
iF_3(q^2)\bar{u}(p_2)\sigma_{\mu\nu}q^\nu \gamma_5 u(p_1).
\end{equation}
The value of this form factor at zero-momentum transfer:
\begin{equation}
F_3(q^2 = 0) \equiv d^\gamma_t,
\end{equation}
is called the EDM. This induces a local interaction that
can be derived from the effective Lagrangian:
\begin{eqnarray}
{\cal L}^d_{eff}=\frac{1}{2}d^\gamma_t\bar{u} 
                     i\sigma_{\mu\nu}\gamma_5 u F^{\mu\nu} 
                 +\frac{1}{2}d^Z_t\bar{u} 
                     i\sigma_{\mu\nu}\gamma_5u Z^{\mu\nu}
                 +\frac{1}{2}d^g_t\bar{u} i\sigma_{\mu\nu}
                     \gamma_5\frac{\lambda_a}{2}u G^{\mu\nu}_a,
\label{eq:edm lagrangian} 
\end{eqnarray}
where we have also added the generalizations to fermion couplings to the 
$Z$ boson and gluons.

The quark EDM in the SM vanishes
at the one-loop order, because of the unitarity of CKM matrix. Diagrams 
at two-loop order can have a $CP$-violating phase, but it has been shown by
Shabalin\cite{Shabalin} that the sum of two-loop contributions to 
$F_3(q^2 = 0)$ vanishes.
It is thus thought that the lowest-order SM contribution
to the quark EDM occurs at least at the three-loop level.

There are models which generate a non-zero quark EDM at 
the one-loop level. Typical examples are the models of 
$CP$ violation with extra scalars\cite{HHG}.
When $CP$ violation comes from the exchange of a neutral Higgs boson, 
the EDM for up-type quarks, down-type quarks, or charged 
leptons is given in the $M_H\gg m_f$ limit by:
\begin{equation}
d^\gamma_f = \frac{eQ_f\sqrt{2}G_F}{32\pi^2}m_f^3
\frac{{\rm Re}(A){\rm Im}(A)}{M_H^2}\log\left(\frac{m_H^2}{M_f^2}\right),
\end{equation}
where $A$ is a dimensionless parameter for the Higgs coupling with the
left-handed fermion.  
This is largest for the top-quark although in the case of the top-quark
it may be a poor approximation to take $q^2 = 0$.
In the case where the $CP$ violation arises in the charged scalar sector, 
the EDM for down-type quarks is given by
\begin{equation}
d^\gamma_d = e\frac{\sqrt{2}G_F}{12\pi^2} m_d {\rm Im}(\alpha_1\beta_1^*)
 |V^*_{td}|\frac{x_t}{(1-x_t)^2}\left( \frac{3}{4} - \frac{5}{4}x_t +
\frac{1 - \frac{3}{2}x_t}{1 - x_t}\log x_t \right),
\end{equation}
where $\alpha_1$ and $\beta_1$ are dimensionless parameters for the 
charged Higgs coupling with fermions, and $x_t = m_t^2/M_H^2$. 
This result follows from the dominance of the
top-quark in the loop and assumes that the dominant contribution comes
from the lightest charged scalar $H^+$. For the case of an up-type 
quark the result is:
\begin{equation}
d^\gamma_u = e\frac{\sqrt{2}G_F}{12\pi^2} m_u {\rm Im}(\alpha_1\beta_1^*)
   \frac{x_t}{(1-x_t)^2} \sum_i |V^*_{ui}|^2
   \left(x_i - \frac{1 - 3x_i}{2(1 - x_i)}\log x_i \right).
\end{equation}

We denote the $\gamma tt$ and $Ztt$ couplings by the vertex 
factor $ie\Gamma^V_\mu$ (See Figure~1), where
\begin{eqnarray}
\Gamma^V_\mu=v_{_V}\gamma_\mu+a_{_V}\gamma_\mu\gamma_5
            +\frac{c_{_V}}{2m_t}\sigma_{\mu\nu}\gamma_5q^\nu,\qquad
            V=\gamma, Z,
\label{eq:vertex}
\end{eqnarray}
with the vector and axial-vector couplings of the $t$ quark given 
in the SM by 
\begin{eqnarray}
v_\gamma=\frac{2}{3},\quad  a_\gamma=0,\quad 
v_{_Z}=\frac{(\frac{1}{4}-\frac{2}{3}x_{_W})}{\sqrt{x_{_W}(1-x_{_W})}},\quad 
a_{_Z}=-\frac{1}{4\sqrt{x_{_W}(1-x_{_W})}},
\end{eqnarray}
and $x_{_W}=\sin^2\theta_W$, $\theta_W$ being the weak mixing angle. 
Here, $q$ is the four-momentum of the vector boson, $V(=\gamma, Z)$. 
For $x_{_W}=0.23$, we find that $v_{_Z}=0.23$ and $a_{_Z}=-0.59$.
We assume that the only additional couplings to the SM ones
are the $CP$-violating EDM and WDM factors, 
\begin{eqnarray}
d^{\gamma,Z}_t=\frac{e}{m_t}c_{\gamma,Z}
            \approx 1.13\times 10^{-16}c_{\gamma,Z}({\rm e}cm),
\end{eqnarray}
for $m_t=175$ GeV, and that the $CP$-violating form factors, $c_{\gamma, Z}$ 
are small.

\section{Electron-positron mode}
\label{sec:Top_EP}

\subsection{Helicity amplitudes for top-quark pair production}

\pr
We define the helicities of the $t$ and $\bar{t}$ in the $e^+e^-$ c.m. 
frame. Let us define the coordinate system $F_0$ for the $t\bar{t}$ 
production process, $e^+e^-\rightarrow t\bar{t}$. The scattering is in 
the $x$-$z$ plane and the $z$-axis is along the top-quark momentum 
direction. The $y$-axis is along $\vec{p}_{e^-}\times\vec{p}_t$ 
and the $x$-axis is given by the right-handed rule. 

We calculate the polarization amplitudes 
${\cal M}_{\sigma,\bar{\sigma};\lambda,\bar{\lambda}} 
=e^2 M_{\sigma,\bar{\sigma};\lambda,\bar{\lambda}}$ for the  production
process $e^+e^-\rightarrow t\bar{t}$ by using a very
straightforward and general method\cite{HZ} based on 
two-component spinors. The helicity amplitudes 
$M_{\sigma,\bar{\sigma};\lambda,\bar{\lambda}}$ are 
presented in the $e^+e^-$ c.m. frame where the positive $z$-axis 
is chosen along the top quark momentum direction :
\begin{eqnarray}
&&p_e=\frac{\sqrt{s}}{2}(1,-\sin\Theta,0,\cos\Theta),\qquad 
  p_{\bar{e}}=\frac{\sqrt{s}}{2}(1,\sin\Theta,0,-\cos\Theta),\nonumber\\
&&p_t=\frac{\sqrt{s}}{2}(1,0,,0,\beta),\qquad\qquad\hskip 0.8cm
  p_{\bar{t}}=\frac{\sqrt{s}}{2}(1,0,0,-\beta).
\label{eq:frame}
\end{eqnarray}
Then the helicity amplitudes of the process $e^+e^-\rightarrow t\bar{t}$
can be expressed in a compact fashion as follows
\begin{eqnarray}
M_{\sigma,\bar{\sigma};\lambda,\bar{\lambda}}&=&
 \frac{1}{s}\sum_{\alpha=L,R}\sum_{\alpha^\prime=\pm}
           (v_\alpha+\alpha^\prime a_\alpha)
           \left[J^e_\alpha(\sigma,\bar{\sigma})\cdot
           J^t_{\alpha^\prime}(\lambda,\bar{\lambda})\right]\nonumber\\
 &&+\frac{i}{2m_ts}\sum_{\alpha=L,R} c_\alpha
              \left[S^e_\alpha(\sigma,\bar{\sigma})\cdot
              P^t(\lambda,\bar{\lambda})\right],
\end{eqnarray}
where
\begin{eqnarray}
&&J^{e\mu}_\alpha(\sigma,\bar{\sigma})=\bar{v}(p_{\bar{e}},\bar{\sigma})
    \gamma^\mu P_\alpha u(p_e,\sigma),\qquad
  J^{t\mu}_{\alpha^\prime}(\lambda,\bar{\lambda})=\bar{u}(p_t,\lambda)
    \gamma^\mu P_{\alpha^\prime} v(p_{\bar{t}},\bar{\lambda}),\nonumber\\
&&S^{e}_{\alpha}(\sigma,\bar{\sigma})=2\bar{v}(p_{\bar{e}},\bar{\sigma})
    \not\! p_t P_\alpha u(p_e,\sigma),\qquad\hskip 0.1cm
  P^{t}(\lambda,\bar{\lambda}) =\bar{u}(p_t,\lambda)
    \gamma_5 v(p_{\bar{t}},\bar{\lambda}),
\end{eqnarray}
with $\sigma,\bar{\sigma}=L,R$, $\lambda,\bar{\lambda}=\pm$,
and $P_{R,L}=P_\pm=\frac{1}{2}\left(1\pm\gamma_5\right)$.

In the vanishing electron mass limit, the positron helicity should 
be opposite to the electron helicity\cite{Kh}, that is to say, 
$M_{LL;\lambda\bar{\lambda}}=M_{RR;\lambda\bar{\lambda}}=0$. 
Therefore, it is convenient to rewrite 
$M_{LR;\lambda\bar{\lambda}}=M^L_{\lambda\bar{\lambda}}$ and
$M_{RL;\lambda\bar{\lambda}}=M^R_{\lambda\bar{\lambda}}$, which are 
given by 
\begin{eqnarray}
M^L_{\mp\pm} =\mp\left(v_{_L}\mp\beta a_{_L}\right)(1\pm\cos\Theta),\quad
M^L_{\mp\mp} =\pm\frac{\sqrt{s}}{2m_t}
              \left[4\frac{m^2_t}{s}v_{_L}\mp i\beta c_{_L}\right]
              \sin\Theta,
\end{eqnarray}
for the initial $e^-_Le^+_R$ configuration, and 
\begin{eqnarray}
M^R_{\mp\pm} =\pm\left(v_{_R}\mp\beta a_{_R}\right)(1\mp\cos\Theta),\quad
M^R_{\mp\mp} =\pm\frac{\sqrt{s}}{2m_t}
              \left[4\frac{m^2_t}{s}v_{_R}\mp i\beta c_{_R}\right]
              \sin\Theta,
\end{eqnarray}
for the initial $e^-_Re^+_L$ configuration
with the scattering angle $\Theta$ and the dimensionless variables 
defined as
\begin{eqnarray}
&&v_{_L}=v_\gamma+r_{_L}v_{_Z},\qquad  v_{_R}=v_\gamma+r_{_R}v_{_Z},\nonumber\\
&&a_{_L}=a_\gamma+r_{_L}a_{_Z},\qquad  a_{_R}=a_\gamma+r_{_R}a_{_Z},\nonumber\\ 
&&c_{_L}=c_\gamma+r_{_L}c_{_Z},\qquad  c_{_R}=c_\gamma+r_{_R}c_{_Z}.
\label{eq:coupling_def1}
\end{eqnarray}
Here the two $\sqrt{s}$-dependent parameters $r_{_L}$ and $r_{_R}$ 
in the SM are defined as
\begin{eqnarray}
r_{_L}&=&\frac{\left(\frac{1}{2}-x_{_W}\right)}{\left(1
   -\frac{m^2_Z}{s}\right)\sqrt{x_{_W}(1-x_{_W})}}
   \approx +0.64\left(1-\frac{m^2_Z}{s}\right)^{-1},\nonumber\\
r_{_R}&=&\frac{-x_{_W}}{\left(1-\frac{m^2_Z}{s}\right)\sqrt{x_{_W}(1-x_{_W})}}
   \approx -0.55\left(1-\frac{m^2_Z}{s}\right)^{-1},
\label{eq:coupling_def2}
\end{eqnarray}
where we have inserted $x_{_W}=0.23$ and have neglected the
$Z$ boson width $\Gamma_{_Z}$, which is easily incorporated but
its numerical effect is minute for 
$\sqrt{s}\geq 2m_t$ since $\Gamma_{_Z}/\sqrt{s}\leq 7\times 10^{-3}$.

\subsection{Top and anti-top quark decays}

\pr
We calculate the helicity amplitudes of  
$t\rightarrow W^+b$ and $\bar{t}\rightarrow W^-\bar{b}$ for 
on-shell $W^\pm$ bosons.
For the process $t\rightarrow W^+b$, the top
quark is taken to decay in its rest frame where the top quark
momentum is $p_t=(m_t,0,0,0)$. Spherical coordinates are 
used to describe the outgoing particles; the polar angle
$\theta$ is taken from the positive $z$ axis and the azimuthal
angle $\phi$ is taken from the positive $x$ axis in the $x$-$y$ plane. 
The bottom quark and $W$ boson are taken on their mass shells with the 
four-momenta $p_b$ ($p_{\bar{b}}$) for the bottom (anti-)quark 
and the four-momenta $p_{_{\bar{W}}}$ ($p_{_W}$) for the $W^+$ ($W^-$)
bosons taken as
\begin{eqnarray}
&&p_b        =E^*_b(1,-\sin\theta\cos\phi,-\sin\theta\sin\phi,-\cos\theta),
              \nonumber\\
&&p_{_{\bar{W}}}=E^*_{\bar{W}}(1,\beta_W\sin\theta\cos\phi,
              \beta_W\sin\theta\sin\phi,\beta_W\cos\theta),\nonumber\\
&&p_{\bar{b}}=E^*_{\bar{b}}(1,-\sin\bar{\theta}\cos\bar{\phi},
             -\sin\bar{\theta}\sin\bar{\phi},-\cos\bar{\theta}),\nonumber\\
&&p_{_W}      =E^*_{W}(1,\beta_W\sin\bar{\theta}\cos\bar{\phi},
              \beta_W\sin\bar{\theta}\sin\bar{\phi},\beta_W\cos\bar{\theta}),
\end{eqnarray}
where we neglect the bottom quark mass, which is about 5 GeV,
and then 
\begin{eqnarray}
E^*_b=\frac{m^2_t-m^2_W}{2m_t},\qquad  
E^*_W=\frac{m^2_t+m^2_W}{2m_t},\qquad \beta^*_W=\frac{E_b}{E_W}.
\end{eqnarray}
The angles $\theta$ ($\bar{\theta}$) and $\phi$ $(\bar{\phi})$ 
in the $t$ ($\bar{t}$) decay refer to the direction of the $W^+$ ($W^-$) boson.
We denote the helicity amplitudes as 
$M^{\bar{W}}_{h_t;\lambda_{\bar{W}},h_b}$
and as $\bar{M}^W_{h_{\bar{t}};\lambda_{W},h_{\bar{b}}}$
after extracting a common factor as
\begin{eqnarray}
&&M^{\bar{W}}_{h_t;\lambda_{\bar{W}},h_b}
          =-\frac{e}{\sqrt{2}\sin\theta_W}
            \sqrt{m^2_t-m^2_W}
            \langle h_t;\lambda_{\bar{W}},h_b\rangle_t,\nonumber\\
&&\bar{M}^W_{h_{\bar{t}};\lambda_{W},h_{\bar{b}}}
            =-\frac{e}{\sqrt{2}\sin\theta_W}
            \sqrt{m^2_t-m^2_W}
           \langle h_{\bar{t}};\lambda_{W},h_{\bar{b}}\rangle_{\bar{t}}.
\end{eqnarray}
There are four non-vanishing helicity amplitudes for each decay mode  
in the rest frame of the top quark and the top anti-quark for 
$m_b=m_{\bar{b}}=0$:
\begin{eqnarray}
&&\langle -;0-\rangle_t=
              \frac{m_t}{m_W}\sin\frac{\theta}{2},\qquad\ \ \ \
  \langle -;--\rangle_t=
              \sqrt{2}\cos\frac{\theta}{2},\nonumber\\
&&\langle +;0-\rangle_t=
              \frac{m_t}{m_W}\cos\frac{\theta}{2}{\rm e}^{i\phi},\qquad
  \langle +;--\rangle_t=
             -\sqrt{2}\sin\frac{\theta}{2}{\rm e}^{i\phi},\\
&&\langle +;0+\rangle_{\bar{t}}=
             -\frac{m_t}{m_W}\cos\frac{\bar{\theta}}{2}
              {\rm e}^{-i\bar{\phi}}, \hskip 0.6cm 
  \langle +;++\rangle_{\bar{t}}=
             -\sqrt{2}\sin\frac{\bar{\theta}}{2}
              {\rm e}^{-i\bar{\phi}},\nonumber\\
&&\langle -;0+\rangle_{\bar{t}}=
             -\frac{m_t}{m_W}\sin\frac{\bar{\theta}}{2},\qquad\hskip 0.6cm 
  \langle -;++\rangle_{\bar{t}}=
             \sqrt{2}\cos\frac{\bar{\theta}}{2}.
\end{eqnarray}
The helicity amplitudes can be used to derive the density matrix of the 
top quark. When the $W$ polarization is not measured,
the $t$ and $\bar{t}$ decay density matrices are given by
\begin{eqnarray}
&&D_t=\frac{1}{2}\left(\begin{array}{cc}
          1+\kappa_{_W}\cos\theta &\ \ \kappa_{_W}\sin\theta {\rm e}^{i\phi}\\
          \kappa_{_W}\sin\theta {\rm e}^{-i\phi} &\ \ 1-\kappa_{_W}\cos\theta
          \end{array}\right),\nonumber\\
&&\bar{D}_{\bar{t}}=\frac{1}{2}\left(\begin{array}{cc}
  1+\kappa_{_W}\cos\bar{\theta} &\ \ \kappa_{_W}\sin\bar{\theta} 
                                   {\rm e}^{-i\bar{\phi}}\\
  \kappa_{_W}\sin\bar{\theta} {\rm e}^{i\bar{\phi}} 
       &\ \ 1-\kappa_{_W}\cos\bar{\theta}
           \end{array}\right),
\end{eqnarray}
respectively, where the polarization efficiency $\kappa_{_W}$ is given by
\begin{eqnarray}
\kappa_{_W}=\frac{m^2_t-2m^2_W}{m^2_t+2m^2_W}\approx 0.41,
\end{eqnarray}
for $m_t=175$ GeV and $m_W=80$ GeV. 

We now discuss the angular distributions of the leptons $l^\pm$
arising from semileptonic decays
\begin{eqnarray}
&&t\rightarrow W^+(p_{\bar{W}})b
   \rightarrow l^+(p_{\bar{l}})\nu b(p_b),\nonumber\\
&&\bar{t}\rightarrow W^-(p_{_W})\bar{b}
         \rightarrow l^-(p_l)\bar{\nu}\bar{b}(p_{\bar{b}}),
\end{eqnarray}
for the polarized $t$ and $\bar{t}$ quarks, where the momenta in 
the parentheses refer to the rest systems of $t$ and $\bar{t}$ and 
serve to analyze the spin polarization of $t$ and $\bar{t}$. 
Neglecting lepton masses we write the lepton momenta as
\begin{eqnarray}
&& p_{\bar{l}}=E^*_{\bar{l}}(1,\sin\theta_l\cos\phi_l,
                    \sin\theta_l\sin\phi_l,\cos\theta_l),\nonumber\\
&& p_l=E^*_{l}(1,\sin\bar{\theta}_l\cos\bar{\phi}_l,
       \sin\bar{\theta}_l\sin\bar{\phi}_l,\cos\bar{\theta}_l),\nonumber\\
&& p_{\nu}=E^*_{\nu}(1,\sin\theta_\nu\cos\phi_\nu,
                \sin\theta_\nu\sin\phi_\nu,\cos\theta_\nu),\nonumber\\
&& p_{\bar{\nu}}=E^*_{\bar{\nu}}
       (1,\sin\bar{\theta}_{\bar{\nu}}\cos\bar{\phi}_{\bar{\nu}},
       \sin\bar{\theta}_{\bar{\nu}}\sin\bar{\phi}_{\bar{\nu}},
       \cos\bar{\theta}_{\bar{\nu}}),
\end{eqnarray}
where $E^*_{\bar{l}}$ and $E^*_l$ are the lepton energies and
$E^*_{\nu}$ and $E^*_{\bar{\nu}}$ the neutrino energies in the
$t$ and $\bar{t}$ rest frames, respectively.
We denote the helicity amplitudes as $M^{\bar{l}}_{h_t}$ and as
$M^l_{h_{\bar{t}}}$ after extracting a common factor as follows
\begin{eqnarray}
&& M^{\bar{l}}_{h_t}=2g^2
   \frac{\sqrt{(m^2_t-q^2)E_\nu E_{\bar{l}}}}{q^2-m^2_W+im_W\Gamma_W}
   \langle h_t,h_b\rangle^{\bar{l}},\nonumber\\   
&& M^l_{h_{\bar{t}}}=2g^2
   \frac{\sqrt{(m^2_t-q^2)E_{\bar{\nu}}E_{l}}}{q^2-m^2_W+im_W\Gamma_W}
   \langle h_{\bar{t}},h_{\bar{b}}\rangle^{l}.
\end{eqnarray}
In the semileptonic decays, there are two non-vanishing helicity amplitudes 
for each decay mode in the rest frames of the top quark and the top 
anti-quark for $m_b=m_{\bar{b}}=0$:
\begin{eqnarray}
&& \langle +,- \rangle^{\bar{l}}=\cos\frac{\theta_l}{2}
   \bigg[\cos\frac{\theta_b}{2}\sin\frac{\theta_\nu}{2}{\rm e}^{i\phi_\nu}
        -\sin\frac{\theta_b}{2}\cos\frac{\theta_\nu}{2}{\rm e}^{i\phi_b}
   \bigg],\nonumber\\
&& \langle -,- \rangle^{\bar{l}}=\sin\frac{\theta_l}{2}{\rm e}^{-i\phi_l}
   \bigg[\cos\frac{\theta_b}{2}\sin\frac{\theta_\nu}{2}{\rm e}^{i\phi_\nu}
        -\sin\frac{\theta_b}{2}\cos\frac{\theta_\nu}{2}{\rm e}^{i\phi_b}
   \bigg],\nonumber\\
&& \langle +,+ \rangle^{l}=-\cos\frac{\bar{\theta}_l}{2}
   \bigg[\cos\frac{\bar{\theta}_b}{2}\sin\frac{\bar{\theta}_\nu}{2}
        {\rm e}^{-i\bar{\phi}_\nu}
        -\sin\frac{\bar{\theta}_b}{2}\cos\frac{\bar{\theta}_\nu}{2}
         {\rm e}^{-i\bar{\phi}_b}
   \bigg],\nonumber\\
&& \langle -,+ \rangle^{l}=-\sin\frac{\bar{\theta}_l}{2}
       {\rm e}^{-i\bar{\phi}_l}
   \bigg[\cos\frac{\bar{\theta}_b}{2}\sin\frac{\bar{\theta}_\nu}{2}
        {\rm e}^{-i\bar{\phi}_\nu}
        -\sin\frac{\bar{\theta}_b}{2}\cos\frac{\bar{\theta}_\nu}{2}
        {\rm e}^{-i\bar{\phi}_b}
   \bigg],
\end{eqnarray}

It is well known that within the SM the angular distribution of the 
charged lepton is a much better spin analyzer of the top quark than that 
of the $b$ quark or the $W$-boson arising from semi- or non-leptonic 
$t$ decays. As a matter of fact, the decay matrices of the semileptonic 
decays of polarized $t$ and $\bar{t}$ are given in the $t$ and $\bar{t}$ 
helicity bases by
\begin{eqnarray}
&&D^{\bar{l}}_t=\frac{1}{2}\left(\begin{array}{cc}
          1+\cos\theta_l &\ \ \sin\theta_l {\rm e}^{i\phi_l}\\
          \sin\theta_l {\rm e}^{-i\phi_l} &\ \ 1-\cos\theta_l
          \end{array}\right),\\
&&\bar{D}^l_{\bar{t}}=\frac{1}{2}\left(\begin{array}{cc}
          1+\cos\bar{\theta}_l &\ \ 
          \sin\bar{\theta}_l {\rm e}^{-i\bar{\phi}_l}\\
          \sin\bar{\theta}_l {\rm e}^{i\bar{\phi}_l} 
         &\ \ 1-\cos\bar{\theta}_l
          \end{array}\right),
\end{eqnarray}
respectively. $\theta_l$ and $\phi_l$ are the polar and azimuthal
angles of $l^+$ from the $t$ decay, which are defined in the $t$ rest frame 
$F_t$ constructed by boosting the $t\bar{t}$ c.m. 
frame $F_0$ along the top quark momentum direction. 
Similarly, the polar angle $\bar{\theta}_l$ and the 
azimuthal angle $\bar{\phi}_l$ of $l^-$ from the $\bar{t}$ decay 
are defined in the $\bar{t}$ rest frame $F_{\bar{t}}$ constructed 
by boosting the $t\bar{t}$ center of mass frame $F_0$ along the anti-top 
quark momentum direction.
Through the present work, it is important to keep in mind that the 
three coordinate systems $F_0$, $F_t$ and $F_{\bar{t}}$ have parallel 
directions of coordinate axes. Note that the polarization efficiency 
is unity in the semileptonic decays, implying that the 
charged lepton analyzes the spin of the top quark much more
efficiently than the corresponding $b$ quark.

To lowest order in the SM and in the narrow-width approximation
we obtain the following normalized distribution
of the semileptonic $t$ decay:
\begin{eqnarray}
&&N(t\rightarrow b\bar{l}\nu)_{\lambda\lambda^\prime}
 =\frac{12x(1-x)}{(1+2w)(1-w)^2}
    \left[D^{\bar{l}}_t(\lambda\lambda^\prime)\right]
    {\rm d}x\frac{{\rm d}\Omega_{\bar{l}}}{4\pi},\\
&&\bar{N}(\bar{t}\rightarrow \bar{b}l\bar{\nu})_{\bar{\lambda}
       \bar{\lambda}^\prime}
 =\frac{12\bar{x}(1-\bar{x})}{(1+2w)(1-w)^2}
    \left[\bar{D}^l_t(\bar{\lambda}\bar{\lambda}^\prime)\right]
    {\rm d}\bar{x}\frac{{\rm d}\Omega_l}{4\pi},
\end{eqnarray}
where $\lambda^{(\prime)}$ and $\bar{\lambda}^{(\prime)}$ refer
to the helicities of the $t$ and $\bar{t}$, respectively, and 
\begin{eqnarray}
&&x=\frac{2E^*_{\bar{l}}}{m_t},\ \ \bar{x}=\frac{2E^*_l}{m_t},\ \
  w=\frac{m^2_W}{m^2_t},\ \ w\leq x(\bar{x}) \leq 1,\nonumber\\
&&{\rm d}\Omega_{\bar{l}}={\rm d}\cos\theta_l{\rm d}\phi_l,\qquad\ \
  {\rm d}\Omega_l={\rm d}\cos\bar{\theta}_l{\rm d}\bar{\phi}_l.
\label{w_definition}
\end{eqnarray}
The factorization of the lepton distribution into an 
energy and angular dependent part holds and this property is 
irrespective of whether the $W$ boson in on-shell or not. 
It was shown in Ref.~\cite{CJK} that even the order $\alpha_s$ QCD 
corrections respect this factorization property to a high degree 
of accuracy.

\subsection{$CP$-odd observables}

\pr
The differential cross section of the process $e^+e^-\rightarrow t\bar{t}$, 
followed by the decays $t\rightarrow bX^+$ and 
$\bar{t}\rightarrow \bar{b}X^-$ is given by
\begin{eqnarray}
&&{\rm d}\sigma\bigg(e^+e^-\rightarrow 
       t\bar{t}\rightarrow bX^+\bar{b}X^-\bigg)_{L,R}
   =\frac{6\pi\alpha^2\beta}{s}
    \bigg[B_{X^+}\bar{B}_{X^-}\bigg]\nonumber\\ 
&&{ } \hskip 1cm\times
 \Sigma_{L,R}(\Theta;\xi_1,\bar{\xi}_1;\xi_2,\bar{\xi}_2;\xi_3,\bar{\xi}_3)
        \bigg[{\rm d}\cos\Theta\bigg]
  \left[\frac{{\rm d}\cos\theta{\rm d}\phi}{4\pi}\right]
  \left[\frac{{\rm d}\cos\bar{\phi}{\rm d}\bar{\phi}}{4\pi}\right],
\end{eqnarray}
where for notational convenience the abbreviations 
\begin{eqnarray}
&&\xi_1=\sin\theta\cos\phi,\qquad 
  \xi_2=\sin\theta\sin\phi,\qquad
  \xi_3=\cos\theta,\nonumber\\
&&\bar{\xi}_1=\sin\bar{\theta}\cos\bar{\phi},\qquad 
  \bar{\xi}_2=\sin\bar{\theta}\sin\bar{\phi},\qquad
  \bar{\xi}_3=\cos\bar{\theta},
\end{eqnarray}
are used, $\Theta$ is the scattering angle between the electron and
top-quark momenta, and $B_{X^+}$ and $\bar{B}_{X^-}$ are the branching 
fractions of $t\rightarrow bX^+$ and $\bar{t}\rightarrow \bar{b}X^-$. 
Here, the angular dependence $\Sigma_{L,R}$ is given by 
\begin{eqnarray}
\Sigma_{L,R}(\Theta;\xi_1,\bar{\xi}_1; \xi_2,\bar{\xi}_2;\xi_3,\bar{\xi}_3)
 \equiv\sum_{\lambda\bar{\lambda}\lambda^\prime\bar{\lambda}^\prime=\pm}
       M^{L,R}_{\lambda\bar{\lambda}}
       M^{*L,R}_{\lambda^\prime\bar{\lambda}^\prime}
       D^X_{\lambda\lambda^\prime}
       \bar{D}^{\bar{X}}_{\bar{\lambda}\bar{\lambda}^\prime}.
\end{eqnarray}
In the $e^+e^-$ c.m frame the angular dependence $\Sigma_{L,R}$ 
for the process $e^+e^-\rightarrow t\bar{t} \rightarrow (X^+b)(X^-\bar{b})$ 
can be written as
\begin{eqnarray}
&&\Sigma_{L,R}(\Theta;\xi_1,\bar{\xi}_1;
  \xi_2,\bar{\xi}_2;\xi_3,\bar{\xi}_3)
  ={\cal P}_{1L,R}{\cal D}_1
  +\kappa\bar{\kappa}{\cal P}_{2L,R}{\cal D}_2\nonumber\\
&&{ } \hskip 0.5cm 
 +\left[\frac{(\kappa+\bar{\kappa})}{2}{\cal P}_{3L,R}
  +\frac{(\kappa-\bar{\kappa})}{2}{\cal P}_{4L,R}\right]{\cal D}_3
 +\left[\frac{(\kappa+\bar{\kappa})}{2}{\cal P}_{4L,R}
  +\frac{(\kappa-\bar{\kappa})}{2}{\cal P}_{3L,R}\right]{\cal D}_4\nonumber\\
&&{ } \hskip 0.5cm
 +\left[\frac{(\kappa+\bar{\kappa})}{2}{\cal P}_{5L,R}
  +\frac{(\kappa-\bar{\kappa})}{2}{\cal P}_{6L,R}\right]{\cal D}_5
 +\left[\frac{(\kappa+\bar{\kappa})}{2}{\cal P}_{6L,R}
  +\frac{(\kappa-\bar{\kappa})}{2}{\cal P}_{5L,R}\right]{\cal D}_6\nonumber\\
&&{ } \hskip 0.5cm +\left[\frac{(\kappa+\bar{\kappa})}{2}{\cal P}_{7L,R}
  +\frac{(\kappa-\bar{\kappa})}{2}{\cal P}_{8L,R}\right]{\cal D}_7
 +\left[\frac{(\kappa+\bar{\kappa})}{2}{\cal P}_{8L,R}
  +\frac{(\kappa-\bar{\kappa})}{2}{\cal P}_{7L,R}\right]{\cal D}_8\nonumber\\
&&{ } \hskip 0.5cm +\kappa\bar{\kappa}
  \left[\sum_{\alpha=9}^{16}{\cal P}_{\alpha L,R}{\cal D}_\alpha\right],
\label{eq:angular_dependence}
\end{eqnarray}
where the definition of the sixteen (16) functions ${\cal P}_{\alpha X}$ 
($X=L,R$) and the sixteen correlation functions ${\cal D}_\alpha$ 
($\alpha=1$ to $16$) is given in Appendicesx~\ref{appendix:p functions} and 
\ref{appendix:cal_D functions}, respectively, and $\kappa(\bar{\kappa})=
\kappa_{_W}$ for the inclusive $t(\bar{t})$ decay and unity for the
semileptonic $t(\bar{t})$ decay.
 
The terms ${\cal P}_\alpha$ and ${\cal D}_\alpha$ can thus be divided 
into four categories under $CP$ and $CP\tilde{T}$: even-even, even-odd, 
odd-even, and odd-odd terms. $CP$-odd coefficients directly measure $CP$ 
violation and $CP\tilde{T}$-odd terms indicate rescattering effects.
Table~1 shows that there exist six (6) independent 
$CP$-odd terms among which ${\cal P}_5$, ${\cal P}_{12}$,
${\cal P}_{14}$ and ${\cal P}_3$ are $CP\tilde{T}$-even, and
${\cal P}_3$, ${\cal P}_{7}$ and ${\cal P}_{16}$ $CP\tilde{T}$-odd.

From now on we make a detailed investigation of the production process 
$e^+e^-\rightarrow t\bar{t}$.  Recently, $CP$ violation in the
production process has been extensively investigated. 
Here, we study the $CP$-violating effects of the process 
$e^+e^-\rightarrow t\bar{t}$ in a rather unified manner. 
As shown previously, we can consider three $CP$-odd and 
$CP\tilde{T}$-even and three $CP$-odd and $CP\tilde{T}$-odd terms. 

Including electron beam polarization, we can obtain 
thirty-two (32) observables in total: 
\begin{eqnarray*}
\begin{array}{ccccc}
32  &  =  &      2                 & \times & (2\times 2)^2 \\
{ } & { } &  \Uparrow              & { } & \Uparrow         \\
{ } & { } & {\rm Electron\ \ Pol.} & { } & 
            t\ \ {\rm and }\ \ \bar{t}\ \  {\rm Pol.}
\end{array}
\end{eqnarray*}
Here, the first 2 is for the electron helicity, and two 2's in the 
parentheses for the degrees of top and anti-top polarizations.  
It is therefore clear that the classification according to the $CP$ and 
$CP\tilde{T}$ transformation properties
gives us a complete set of observables that can be measured in the  
process $e^+e^-\rightarrow t\bar{t}$ with left- and right-handed polarized 
electron beams.

The $CP$-odd part of the angular dependence 
(\ref{eq:angular_dependence}) can be separated into two parts 
\begin{eqnarray}
&&\Sigma^{\rm CP}_{L,R}(\Theta;\xi_1,\bar{\xi}_1;
                        \xi_2,\bar{\xi}_2;\xi_3,\bar{\xi}_3)
   =\Sigma^{\rm CP}_{{\rm E}L,R}+\Sigma^{\rm CP}_{{\rm O}L,R},
\end{eqnarray}
where $\Sigma^{\rm CP}_{{\rm E}L,R}$ and $\Sigma^{\rm CP}_{{\rm O}L,R}$
terms are $CP\tilde{T}$-even and $CP\tilde{T}$-odd, respectively, 
and given by
\begin{eqnarray}
&& \!\!\!\Sigma^{\rm CP}_{{\rm E}L,R}
  =
{\cal P}_{5L,R}\left[\frac{(\kappa+\bar{\kappa})}{2}{\cal D}_5
                        +\frac{(\kappa-\bar{\kappa})}{2}{\cal D}_6\right]
    +\kappa\bar{\kappa}\left[{\cal P}_{12L,R}{\cal D}_{12}
                        +{\cal P}_{14L,R}{\cal D}_{14}\right],\\
&& \!\!\!\Sigma^{\rm CP}_{{\rm O}L,R}
  =
 {\cal P}_{3L,R}\left[\frac{(\kappa+\bar{\kappa})}{2}{\cal D}_3
                      +\frac{(\kappa-\bar{\kappa})}{2}{\cal D}_4\right]
 +{\cal P}_{7L,R}\left[\frac{(\kappa+\bar{\kappa})}{2}{\cal D}_7
                      +\frac{(\kappa-\bar{\kappa})}{2}{\cal D}_8\right]
  \nonumber\\
 &&\hskip 1.2cm {} +\kappa\bar{\kappa}{\cal P}_{16}{\cal D}_{16}.
\label{eq:square}
\end{eqnarray}
The explicit form of all the $CP$-odd terms is listed in 
Appendix~\ref{appendix:p functions}.
The upper three $CP$-odd ${\cal P}_{\alpha X}$ terms are 
$CP\tilde{T}$-even and the lower three $CP$-odd terms are 
$CP\tilde{T}$-odd. 

Including electron beam polarization, Poulose and Rindani\cite{PR}  
recently have considered two new $CP$-odd and $CP\tilde{T}$-even 
asymmetries, which are essentially equivalent to the so-called 
triple vector products, and two new $CP$-odd and $CP\tilde{T}$-odd 
asymmetries in addition to the two conventional lepton energy asymmetries. 
It is clear that we can use six more asymmetries among which four 
asymmetries are $CP$-odd and $CP\tilde{T}$-even and the other two
terms are $CP$-odd and $CP\tilde{T}$-odd.

\subsection{Top-quark momentum reconstruction}
\label{subsec:Reconstruction}

\pr
Purely semileptonic decay modes of a $t\bar{t}$ pair also
give the cleanest signal for the top-pair production 
process in $e^+e^-$ collisions:
\begin{eqnarray}
e^-(p_e)+e^+(p_{\bar{e}})&\rightarrow& t(p_t) + \bar{t}(p_{\bar{t}}),
        \nonumber\\
t(p_t)&\rightarrow& b(p_b)+W^+(p_{_{\bar{W}}}),\nonumber\\
\bar{t}(p_{\bar{t}})&\rightarrow&\bar{b}(p_{\bar{b}})+W^-(p_{_W}),
     \nonumber\\
&&\mbox{ } \hskip 1.2cm 
  W^+(p_{_{\bar{W}}})\rightarrow \bar{l}(p_{\bar{l}})+\nu(p_{\nu}),
     \nonumber\\
&&\mbox{ } \hskip 1.25cm
  W^-(p_{_W})\rightarrow l(p_l)+ \bar{\nu}(p_{\bar{\nu}}).
\label{eq:semileptonic}
\end{eqnarray}
The process is observed experimentally as shown in Figure~2;
\begin{eqnarray}
e^++e^-\rightarrow b+\bar{b}+l+\bar{l}
                    +{\rm missing\ \ momentum},
\end{eqnarray}
where the final lepton pair can be either one of $e^-e^+$, 
$e^-\mu^+$, $e^-\tau^+$, $\mu^-e^+$, and $\mu^-\mu^+$.  
The four-momenta of the particles are given in parentheses. 

A simple kinematical analysis, presented below, shows that
the two unobserved neutrino momenta can be determined from the 
observed $b$, $\bar{b}$, and lepton momenta with no ambiguity,
in the limit where the $t$ and $W$ widths and photon (or gluon) 
radiation are neglected. 

The kinematics of the process (\ref{eq:semileptonic}) is determined
by ten angles, two for the scattering,  four each for the semileptonic
$t$ decays. Since we observe the four three-momenta of the final
particles, generally we have superfluous observables to fix the
whole configuration. Here we present an explicit solution
for the two momenta $p_{_W}$ and $p_{_{\bar{W}}}$ in terms of the 
observed $b$, $\bar{b}$, and lepton momenta so that the $t$ and 
$\bar{t}$ momenta can be reconstructed.

It suffices to solve for the three-momentum $\vec{p}_{\bar{W}}$
and then $p_{_W}$ is given by momentum conservation. 
As the $t$ energy is equal to the beam energy $E$, we have
\begin{eqnarray}
p^0_{_{\bar{W}}}=E-p^0_{b}, \qquad
\vec{p}^2_{_{\bar{W}}}=(E-p^0_{b})^2-m^2_W.
\label{eq:eq1}
\end{eqnarray}
A similar equation holds for the $\bar{t}\rightarrow \bar{b}W^-$ decay:
\begin{eqnarray}
\vec{p}^2_{_W}=(E-p^0_{\bar{b}})^2-m^2_W.
\label{eq:eq2}
\end{eqnarray}
Using momentum conservation 
$\vec{p}_{_W}=-\vec{p}_{_{\bar{W}}}+\vec{p}_b+\vec{p}_{\bar{b}}$ and 
Eq.~(\ref{eq:eq1}), this last equation can be written in terms of
$\vec{p}_{_{\bar{W}}}$:
\begin{eqnarray}
(\vec{p}_b+\vec{p}_{\bar{b}})\cdot\vec{p}_{_{\bar{W}}}
   =
E(p^0_b-p^0_{\bar{b}})-p^{02}_b
       -\vec{p}_b\cdot\vec{p}_{\bar{b}}+m^2_b.
\label{eq:eq3}
\end{eqnarray}
The third constraint comes from the condition that the $b$-$W^+$ 
system should have the mass of the $t$ quark:
\begin{eqnarray}
(p_b+p_{_{\bar{W}}})^2=m^2_t,
\end{eqnarray}
which gives
\begin{eqnarray}
\vec{p}_b\cdot\vec{p}_{_{\bar{W}}}=Ep^0_b-p^{02}_b
          +\frac{1}{2}(m^2_W+m^2_b-m^2_t).
\label{eq:eq4}
\end{eqnarray}
Eqs.~(\ref{eq:eq3}) and (\ref{eq:eq4}) lead to
\begin{eqnarray}
\vec{p}_{\bar{b}}\cdot\vec{p}_{_{\bar{W}}}
    =-Ep^0_{\bar{b}}-\vec{p}_b\cdot\vec{p}_{\bar{b}}
     +\frac{1}{2}(m^2_t+m^2_b-m^2_W).
\label{eq:eq5}
\end{eqnarray}

The sequential $W^+\rightarrow \bar{l}\nu$ decay yields another
condition:
\begin{eqnarray}
(p_{_{\bar{W}}}-p_{\bar{l}})^2=0,
\end{eqnarray}
which gives
\begin{eqnarray}
\vec{p}_{\bar{l}}\cdot\vec{p}_{_{\bar{W}}}
   =
 Ep^0_{\bar{l}}-p^0_bp^0_{\bar{l}}-\frac{1}{2}(m^2_W+m^2_{\bar{l}}).
\label{eq:eq6}
\end{eqnarray}

The four conditions (\ref{eq:eq1}), (\ref{eq:eq4}), (\ref{eq:eq5}), and
(\ref{eq:eq6}) provide the solution for $\vec{p}_{\bar{W}}$. 
We rewrite the right-hand sides of these equations for the sake 
of clarity:
\begin{eqnarray}
\vec{p}^2_{_{\bar{W}}}=K,\qquad
\vec{p}_b\cdot\vec{p}_{_{\bar{W}}}=L,\qquad
\vec{p}_{\bar{b}}\cdot\vec{p}_{_{\bar{W}}}=M,\qquad
\vec{p}_{\bar{l}}\cdot\vec{p}_{_{\bar{W}}}=N.
\label{eq:conditions}
\end{eqnarray}

Let us assume, for the moment, that the three three-momenta $\vec{p}_b$, 
$\vec{p}_{\bar{b}}$ and $\vec{p}_{\bar{l}}$ are not parallel. 
Then we can expand $\vec{p}_{_{\bar{W}}}$ in terms of any combination of
two momenta among the three momenta. Here, we choose $\vec{p}_b$ and
$\vec{p}_{\bar{b}}$, for which $\vec{p}_{_{\bar{W}}}$ is expressed as
\begin{eqnarray}
\vec{p}_{_{\bar{W}}}=a\vec{p}_b+b\vec{p}_{\bar{b}}
                 +c\vec{p}_b\times\vec{p}_{\bar{b}}.
\end{eqnarray}

The second and third expressions in Eq.~(\ref{eq:conditions}) constrain
$\vec{p}_{_{\bar{W}}}$ to lie on a line in three-dimensional space. 
They give 
\begin{eqnarray}
&&a\vec{p}^2_b+b\vec{p}_b\cdot\vec{p}_{\bar{b}}=L,\nonumber\\
&&a\vec{p}_b\cdot\vec{p}_{\bar{b}}+b\vec{p}^2_{\bar{b}}=M,
\end{eqnarray}
which can be explicitly solved:
\begin{eqnarray}
\left(\begin{array}{c}
      a \\
      b\end{array}
\right)
  =
\frac{1}{|\vec{p}_b\times\vec{p}_{\bar{b}}|^2}
\left(\begin{array}{cc}
      \vec{p}^2_{\bar{b}} & -\vec{p}_b\cdot\vec{p}_{\bar{b}}\\
     -\vec{p}_b\cdot\vec{p}_{\bar{b}} & \vec{p}^2_b
      \end{array}
\right)
\left(\begin{array}{c}
       L \\
       M
      \end{array}
\right).
\end{eqnarray}
The remaining variable $c$ is determined using the final two 
conditions of Eq.~(\ref{eq:conditions}):
\begin{eqnarray}
c^2&=&\frac{1}{|\vec{p}_b\times\vec{p}_{\bar{b}}|^2}
           \bigg[K -a^2\vec{p}^2_b-b^2\vec{p}^2_{\bar{b}}
                  -2ab\vec{p}_b\cdot\vec{p}_{\bar{b}}\bigg],\\
c  &=&\frac{1}{\vec{p}_{\bar{l}}
                    \cdot(\vec{p}_b\times\vec{p}_{\bar{b}})}
           \bigg[N -a\vec{p}_{\bar{l}}\cdot\vec{p}_b
                  -b \vec{p}_{\bar{l}}\cdot\vec{p}_{\bar{b}}\bigg].
\end{eqnarray}
The sign of $c$ can not be determined by the first equation, but this 
twofold discrete ambiguity is cleared out through
the second constraint which stems from the extra information on 
the antilepton momentum. 

There are two exceptional cases where the $t$ and $\bar{t}$ momenta can not
be determined. 
(i) In the exceptional case that two momenta are parallel, 
      one has a twofold discrete ambiguity, and 
(ii) in the more exceptional case that three momenta are parallel, 
      one obtains an one-parameter family of solution for which the 
      azimuthal angle of $\vec{p}_{_{\bar{W}}}$ with respect to $\vec{P}_b$ 
      is left undetermined.
Even from experimental point of view such two cases are so exceptional
that the reconstruction of the $t$ and $\bar{t}$ momenta can be almost
always possible.

\subsection{$CP$-odd observables in the laboratory frame}

\pr
Experimentally it is, however, difficult to perform a 16-parameter
fit (corresponding to the 16 angular coefficients) for each of
several $\cos\Theta$ bins. Rather one would like to obtain from
the experimental data the moments of those angular distributions
that are most sensitive to new physics, i.e. the anomalous
$CP$-violating form factors, $c_\gamma$ and $c_{_Z}$ at hand.  
However, a sufficiently precise reconstruction of the top quark
direction is required to measure all the angular variables. 
The reconstruction is easy if either the top quark or the
top anti-quark decays into a $b$ quark and $W$ boson that 
decays hadronically. We have shown in Section~\ref{subsec:Reconstruction}
that in the process $e^+e^-\rightarrow t\bar{t}$ even the purely 
semileptonic decays of the $t$ and $\bar{t}$ quarks allow  
the full reconstruction of the particle momenta, especially the top 
and anti-top momenta.
In practice, the use of the directly measurable momenta of the
charged leptons and/or $b$-jets might be easier.
When the top quark and anti-top quark directions are not determined,
the cross section should be rewritten in the laboratory frame and 
variables independent of the top quark and anti-quark directions
should be used. These transformations can be straightforwardly  
performed and several useful angular variables can be introduced.

The previous works have concentrated on the $CP$-odd observables
expressed in terms of the directly measurable particle momenta.
However, the analytic expressions of those observables are very
much involved even if the $CP$-odd terms for the 
specific $t$ and $\bar{t}$ helicity values are very simple.  

First of all we investigate the kinematics of the production-decay 
sequence
\begin{eqnarray}
e^-e^+\rightarrow t\bar{t}\rightarrow X^+b X^-\bar{b}.
\end{eqnarray}
The $t$ and $\bar{t}$ momenta are, of course, back to back. 
The $b$ and $\bar{b}$ momenta can be measured. 
In the laboratory frame, the momenta of $X^+(q_X,\vartheta, \varphi)$ and 
$X^-(q_{\bar{X}},\bar{\vartheta},\bar{\varphi})$ 
are referred with respect to the direction of the top quark. 
The boosts between the
laboratory frame and each of the top and anti-top rest frames are 
defined by the parameters $\gamma=\sqrt{s}/(2m_t)$ and 
$\beta=\sqrt{1-\gamma^{-2}}$.
The momentum variables between the laboratory frame and the top rest frame
are related by
\begin{eqnarray}
E_{\bar{X}}=\gamma(E^*_{\bar{X}}+\beta q^*_{\bar{X}}\cos\theta),&&\qquad
\varphi=\phi,\nonumber\\
q_{\bar{X}}\cos\vartheta=\gamma(q^*_{\bar{X}}\cos\theta
       +\beta E^*_{\bar{X}}),&&\qquad
q_{\bar{X}}\sin\vartheta=q^*_{\bar{X}}\sin\theta,
\end{eqnarray}
and those between the laboratory frame and the anti-top rest frame are
related by 
\begin{eqnarray}
E_X=\gamma(E^*_X-\beta q^*_X\cos\bar{\theta}),&&\qquad
\bar{\varphi}=\bar{\phi},\nonumber\\
q_X\cos\bar{\vartheta}=\gamma(q^*_X\cos\bar{\theta}
              -\beta E^*_X),&&\qquad
q_X\sin\bar{\vartheta}=q^*_X\sin\bar{\theta}.
\end{eqnarray}

Observables which are constructed from the (unit) momenta of the
charged leptons and/or $b$ jets originating from $t$ and $\bar{t}$
decay are directly measurable in future experiments. 
Both the nonleptonic and semileptonic decay channels 
\begin{eqnarray}
&&t\rightarrow bX_{\rm had},\\
&&t\rightarrow bl^+\nu;\ \ l=e,\mu,\tau,
\end{eqnarray}
together with the corresponding charge-conjugated ones are
used in the following. The first set of observables which we consider
involves the momentum of a lepton or $b$ jet from $t$ decay 
correlated with the momentum of a lepton or $b$ jet from
$\bar{t}$ decay. These correlations apply to the reactions
\begin{eqnarray}
e^+(\vec{p}_{\bar{e}})+e^-(\vec{p}_{e})
   \rightarrow t+\bar{t}
   \rightarrow a(\vec{q}_+)+\bar{c}(\vec{q}_-)+X,
\label{eq:inclusive}
\end{eqnarray}
where we use the notation $a,c = e^+,\mu^+,\tau^+,b$ jet, and
$\bar{a},\bar{c}$ will denote the corresponding charge conjugate
particles. The momenta $\vec{p}_{e,\bar{e}}$ and $\vec{q}_\pm$ 
are defined in the overall c.m. frame. Light quark jets resulting
from the hadronic decays are difficult to identify and are
therefore not used for constructing observables in the following.
We shall assume that the $\tau$ momentum is measurable with
a suitable vertex chamber. The subsequent analysis holds for all 
reactions of the form irrespectively of the intermediate $t\bar{t}$
state and of the unobserved part $X$ of the final state.

Let us start with the $CP$-odd energy asymmetries
\begin{eqnarray}
A^b_E=E_b-E_{\bar{b}},\qquad
A^l_E=E_{\bar{l}}-E_l.
\end{eqnarray}
The $CP$-odd asymmetries are proportional to the
$CP\tilde{T}$-odd correlation function ${\cal D}_7$:
\begin{eqnarray}
A^b_E=-\sqrt{\frac{2}{3}}\gamma\beta E^*_b{\cal D}_7,\qquad
\langle A^l_E\rangle_E =\sqrt{\frac{2}{3}}\gamma\beta 
\langle E^*_l\rangle_E {\cal D}_7,
\end{eqnarray}
where $E^*_b=(m^2_t-m^2_W)/2m_t$ and the notation $\langle X \rangle_E$ 
denotes the average of the observable $X$ over the lepton energy 
distribution. Explicitly we obtain for the average of the lepton energy
\begin{eqnarray}
\langle E^*_l \rangle_E
  =\frac{m_t}{2}\int^1_w{\rm d}x\frac{6x^2(1-x)}{(1+2w)(1-w)^2}
  =\frac{m_t}{4}\left[\frac{1+2w+3w^2}{1+2w}\right],
\end{eqnarray}
with $w$ defined in Equation~(\ref{w_definition}),
and for the expectations of the $CP$-odd energy asymmetries
\begin{eqnarray}
&& \langle A^b_E\rangle_{L,R}
         =\frac{4}{9}E^*_b\beta^2\gamma\kappa_{_W} 
          v_{L,R}{\rm Im}(c_{L,R}),\nonumber\\
&& \langle\langle A^b_E\rangle_E\rangle_{L,R}
         =-\frac{4}{9}\langle E^*_l\rangle_E\beta^2\gamma 
          v_{L,R}{\rm Im}(c_{L,R}).
\end{eqnarray}

Secondly, let us investigate the $CP$-odd vector observables
\begin{eqnarray}
&&A^b_1=\hat{p_e}\cdot (\vec{p}_b\times\vec{p}_{\bar{b}}),\qquad
  A^b_2=\hat{p_e}\cdot (\vec{p}_b+\vec{p}_{\bar{b}}),\\
&&A^l_1=\hat{p_e}\cdot (\vec{p}_{\bar{l}}\times\vec{p}_l),\qquad
  A^l_2=\hat{p_e}\cdot (\vec{p}_{\bar{l}}+\vec{p}_l).
\end{eqnarray}
The four $CP$-odd vector observables can be expressed 
in the $t$ and $\bar{t}$ rest frames in terms of the
angular correlations ${\cal D}_\alpha$ ($\alpha=1$ to $16$)
defined in Appendix~\ref{appendix:cal_D functions}, as 
\begin{eqnarray}
&&A^b_1=\frac{\sqrt{2}}{3}E^{*2}_b
         \bigg[\cos\Theta {\cal D}_{12}-\gamma\sin\Theta{\cal D}_{14}
              -\sqrt{3}\gamma\beta\sin\Theta {\cal D}_5\bigg],\nonumber\\
&&A^b_2=-\sqrt{\frac{2}{3}}E^*_b
         \bigg[\gamma\cos\Theta {\cal D}_7-\sin\Theta{\cal D}_3\bigg],\\
&&\langle A^l_1\rangle_E
  =\frac{\sqrt{2}}{3}\langle E^*_l\rangle^2_E
         \bigg[\cos\Theta {\cal D}_{12}-\gamma\sin\Theta{\cal D}_{14}
              +\sqrt{3}\gamma\beta\sin\Theta {\cal D}_5\bigg],\nonumber\\
&&\langle A^l_2\rangle_E 
  =\sqrt{\frac{2}{3}}\langle E^*_l\rangle_E
         \bigg[\gamma\cos\Theta {\cal D}_7-\sin\Theta{\cal D}_3\bigg].
\end{eqnarray}

It is now straightforward to obtain the analytic expressions for 
the the expectation of the observables 
$O^X_i(=A^b_i, \langle A^l_i\rangle_E$) ($X=b,l$ and $i=1,2$) 
defined by
\begin{eqnarray}
\langle O^X_i\rangle=\frac{1}{2}\frac{1}{(4\pi)^2}
        \int^1_{-1}{\rm d}\cos\Theta
        \int {\rm d}\Omega\int {\rm d}\bar{\Omega}
        \bigg[O^X_i 
        \Sigma(\Theta;\xi_1,\bar{\xi}_1;
               \xi_2,\bar{\xi}_2;\xi_3,\bar{\xi}_3)
        \bigg],
\end{eqnarray}
where ${\rm d}\Omega={\rm d}\cos\theta{\rm d}\phi$ and
${\rm d}\bar{\Omega}={\rm d}\cos\bar{\theta}{\rm d}\bar{\phi}$.
We obtain for the expectations of the $CP$-odd vector observables 
\begin{eqnarray}
&&\langle A^b_1\rangle_{L,R}
 =\pm\frac{4}{27}E^{*2}_b\beta\gamma^2\kappa_{_W}
  \bigg(3v_{L,R}-\kappa_{_W} a_{L,R}\bigg){\rm Re}(c_{L,R}),\nonumber\\
&&\langle A^b_2\rangle_{L,R}
 =\pm\frac{4}{9}E^*_b\beta^2\gamma\kappa_{_W} a_{L,R}{\rm Im}(c_{L,R}),
  \nonumber\\
&&\langle \langle A^l_1 \rangle_E \rangle_{L,R}
 =\mp\frac{4}{27}\langle E^*_l\rangle^2_E\beta\gamma^2\
  \bigg(3v_{L,R}+a_{L,R}\bigg){\rm Re}(c_{L,R}),\nonumber\\
&&\langle \langle A^l_2\rangle_E \rangle_{L,R}
 =\mp\frac{4}{9}\langle E^*_l \rangle_E\beta^2\gamma\ 
  a_{L,R}{\rm Im}(c_{L,R}).
\label{eq:cuypers1}
\end{eqnarray}

Thirdly, we turn to the $CP$-odd tensor observables and take $i,j=3$
to consider the $(3,3)$ components with respect to the electron
momentum direction
\begin{eqnarray}
&&T^b_{33}=2(\vec{p}_b-\vec{p}_{\bar{b}})_3
            (\vec{p}_b\times\vec{p}_{\bar{b}})_3,\nonumber\\
&&Q^b_{33}=2(\vec{p}_b+\vec{p}_{\bar{b}})_3
            (\vec{p}_b-\vec{p}_{\bar{b}})_3
           -\frac{2}{3}(\vec{p}^2_b-\vec{p}^2_{\bar{b}}),\\
&&T^l_{33}=2(\vec{p}_{\bar{l}}-\vec{p}_l)_3
            (\vec{p}_{\bar{l}}\times\vec{p}_l)_3,\nonumber\\
&&Q^l_{33}=2(\vec{p}_{\bar{l}}+\vec{p}_l)_3
            (\vec{p}_{\bar{l}}-\vec{p}_l)_3
           -\frac{2}{3}(\vec{p}^2_{\bar{l}}-\vec{p}^2_l).
\end{eqnarray}
The four $CP$-odd tensor observables 
can be expressed in the $t$ and $\bar{t}$ rest frames 
in terms of the angular correlations ${\cal D}_\alpha$ ($\alpha=1$ to $16$)
and some extra angular correlations ${\cal D}^\prime_\beta$ 
($\beta=1$ to $12$), which are defined in 
Appendix~\ref{appendix:cal_D functions}, as
\begin{eqnarray}
T^b_{33}&=&\frac{2\sqrt{2}}{9}E^{*3}_b
  \bigg[-\{7\sqrt{3}(\gamma^2-1){\cal D}_5
      -9\gamma^2\beta{\cal D}_{14}\}\sin\Theta\cos\Theta\     
      \nonumber \\
 && {}\hskip 1cm +3\gamma\beta(3\cos^2\Theta-1){\cal D}_{12}\bigg]\nonumber\\
  &&+\frac{2\sqrt{10}}{15}E^{*3}_b
  \bigg[(\sqrt{3}\gamma^2\beta{\cal D}^\prime_{4}+{\cal D}^\prime_{6}
      +\frac{\sqrt{3}}{3}(2\gamma^2+1){\cal D}^\prime_{8}\nonumber \\
 && { }\hskip 1cm -{\cal D}^\prime_{7}+\gamma^2{\cal D}^\prime_{9})
    \sin\Theta\cos\Theta
    -\gamma({\cal D}^\prime_{11}+\sqrt{3}\beta{\cal D}^\prime_{3})\sin^2\Theta 
      \nonumber\\
 &&{ } \hskip 1cm +\gamma\cos^2\Theta{\cal D}^\prime_{10}
    -\gamma(2\cos^2\Theta-1){\cal D}^\prime_{11}\bigg],\nonumber \\ 
Q^b_{33}&=&\frac{4\sqrt{6}}{9}E^{*2}_b\gamma\beta
  \bigg[3\sin\Theta\cos\Theta{\cal D}_3
      -\gamma(3\cos^2\Theta-1){\cal D}_7\bigg]\nonumber\\
  &&-\frac{2\sqrt{10}}{15}E^{*2}_b
  \bigg[\frac{1}{3}(2\gamma^2+1)(3\cos^2\Theta-1){\cal D}^\prime_{1}
      -\sqrt{3}\sin^2\Theta{\cal D}^\prime_{2}
     \nonumber \\
  &&{ } \hskip 1cm  +2\sqrt{3}\gamma\sin\Theta\cos\Theta{\cal D}^\prime_{5}\bigg],\\
\langle T^l_{33}\rangle_E&=&\frac{2\sqrt{2}}{9}
  \langle E^{*2}_l\rangle_2\langle E^*_l\rangle_E
  \bigg[\{7\sqrt{3}(\gamma^2-1){\cal D}_5
      -9\gamma^2\beta{\cal D}_{14}\}\sin\Theta\cos\Theta\     
      \nonumber \\
  &&{ } \hskip 1cm +3\gamma\beta(3\cos^2\Theta-1){\cal D}_{12}\bigg]\nonumber\\
  &&+\frac{2\sqrt{10}}{15}
  \langle E^{*2}_l\rangle_2\langle E^*_l\rangle_E
  \bigg[(\sqrt{3}\gamma^2\beta{\cal D}^\prime_{4}-{\cal D}^\prime_{6}
      -\frac{\sqrt{3}}{3}(2\gamma^2+1){\cal D}^\prime_{8}
  \nonumber \\
  &&{ } \hskip 1cm +{\cal D}^\prime_{7}-\gamma^2{\cal D}^\prime_{9})
        \sin\Theta\cos\Theta
     +\gamma({\cal D}^\prime_{12}-\sqrt{3}\beta{\cal D}^\prime_{3})\sin^2\Theta
      \nonumber\\
 && { } \hskip 1cm-\gamma\cos^2\Theta{\cal D}^\prime_{10}
    +\gamma(2\cos^2\Theta-1){\cal D}^\prime_{11}\bigg],\nonumber \\ 
\langle Q^l_{33}\rangle_E&=&-\frac{4\sqrt{6}}{9}
  \langle E^{*2}_l\rangle_E \gamma\beta
  \bigg[3\sin\Theta\cos\Theta{\cal D}_3
      -\gamma(3\cos^2\Theta-1){\cal D}_7\bigg]\nonumber\\
  &&-\frac{2\sqrt{10}}{15}\langle E^{*2}_l\rangle_E
  \bigg[\frac{1}{3}(2\gamma^2+1)(3\cos^2\Theta-1){\cal D}^\prime_{1}
      -\sqrt{3}\sin^2\Theta{\cal D}^\prime_{2}\nonumber\\
  &&{ } \hskip 1cm+2\sqrt{3}\gamma\sin\Theta\cos\Theta{\cal D}^\prime_{5}\bigg],
\end{eqnarray}
where the average of the lepton energy squared $\langle E^{*2}_l\rangle$
is given by
\begin{eqnarray}
\langle E^{*2}_l\rangle_E=\frac{m^2_t}{4}\int^1_w{\rm d}x
    \frac{6x^3(1-x)}{(1+2w)(1-w)^2}=\frac{3m^2_t}{40}
    \left[\frac{1+2w+3w^2+4w^3}{1+2w}\right],
\end{eqnarray}
with $w=m^2_W/m^2_t$.
We do not present the analytic expressions for the expectation
values of the $CP$-odd tensor observables although they are
straightforward to obtain.

\subsection{Observable consequences of the top-quark EDM}

\pr
For the sake of numerical analysis we insert the values of the SM vector 
and axial-vector couplings  and then we obtain
\begin{eqnarray}
&&v_{_L}=0.67+0.15\delta_Z,\qquad
  v_{_R}=0.67-0.13\delta_Z,\nonumber\\
&&a_{_L}=-0.38\delta_Z,\qquad
  \hskip 0.8cm a_{_R}=0.32\delta_Z,\nonumber\\
&&c_{_L}=c_\gamma+0.64\delta_Z c_{_Z},\qquad
  c_{_R}=c_\gamma-0.55\delta_Z c_{_Z},
\label{eq:coupling_num}
\end{eqnarray}
where $\delta_Z=(1-m^2_Z/s)^{-1}$. 
The contribution from the $Z$-boson exchange diagram decreases
as the c.m. energy $\sqrt{s}$ increases. 
For $m_t=175$ GeV and $m_Z=91.2$ GeV, $1\leq \delta_Z\leq 1.073$.
Note that the $c_{_Z}$ contribution to $c_{_L}$ and $c_{_R}$ is similar in size 
but different in sign. 
Naturally, the electron polarization is expected to play
a crucial role in discriminating $c_\gamma$ and $c_{_Z}$, as pointed 
out earlier by Cuypers and Rindani\cite{CR}. 

If a non-vanishing expectation value $\langle O_{X}\rangle$
for a given observable $O_{X}$ is observed, it has a statistical 
significance as far as it is compared with the expectation 
variance $\langle O^2_{X}\rangle$.
For instance, to observe a deviation from the SM expectation
with better than one-standard deviations one needs
\begin{eqnarray}
\langle O_{X}\rangle\geq
        \sqrt{\frac{\langle O_{X}^2\rangle}{N_{t\bar{t}}}},\qquad 
N_{t\bar{t}}=\varepsilon\left[B_{X^+}\bar{B}_{X^-}\right]
        {\cal L}_{ee}\sigma(e^+e^-\rightarrow t\bar{t}),
\label{eq:deviation;top}
\end{eqnarray}
where $N_{t\bar{t}}$ is the number of events,  ${\cal L}_{ee}$ 
is the $e^+e^-$ collider luminosity, and $\varepsilon$ is the
detection efficiency.

Implementing Eq.~(\ref{eq:deviation;top}) we can determine the areas 
in the $(c_\gamma,c_{_Z})$ plane which can not be explored with a given  
confidence level. Clearly, because of the linear dependence of the 
expectation values on the $CP$-odd electroweak dipole form factors,
these areas are delimited by straight lines which are equidistant 
from the SM expectation $c_\gamma=c_{_Z}=0$. 
The slopes of these straight lines vary with the polarization degree 
of the initial $e^+e^-$ beams. The use of more than two $CP$-odd 
distributions can help to determine independently the real and 
imaginary parts of the electric as well as weak dipole couplings. 
Of course, longitudinal beam polarization, if present, obviates 
the need for the simultaneous measurement of more than one distribution 
and it can enhance the sensitivity to the $CP$-odd parameters.
Our numerical results are presented for the assumed detection efficiency
$\varepsilon=10\%$ and for the following set of experimental
parameters:
\begin{eqnarray}
&&\sqrt{s}=0.5\ \ {\rm TeV},\qquad
  L_{ee}= \left\{
  \begin{array}{ll}
  10 \ {\rm fb}^{-1} & \ \ \ {\rm for\ \ polarized\ \ electrons},\\ 
  20 \ {\rm fb}^{-1} & \ \ \ {\rm for\ \ unpolarized\ \ electrons},
  \end{array} \right.
\end{eqnarray}

The shadowed parts in Figure~3 show the 1-$\sigma$ allowed 
regions of the $CP$-odd parameters ${\rm Re}(c_\gamma$) and ${\rm Re}(c_{_Z})$ 
through the $CP$-odd and $CP\tilde{T}$-even asymmetries (a) $A^b_{1}$ 
and $T^b_{33}$ and (b) $A^l_{1}$ and $T^l_{33}$ with left- and right-handed
polarized electron beams, respectively. 
The solid lines with a positive (negative) slope are for 
$A^b_{1}$ and $A^l_{1}$ with right-handed (left-handed) electrons while 
the long-dashed lines with a positive (negative) slope are for $T^b_{33}$ 
and $T^l_{33}$ with right-handed (left-handed) electrons.
On the other hand, the shadowed parts in Figure~4 show 
the 1-$\sigma$ allowed regions of the parameters ${\rm Re}(c_\gamma)$ 
and ${\rm Re}(c_{_Z})$ through (a) $A^b_{1}$ and 
$T^b_{33}$ and (b) $A^l_{1}$ and $T^l_{33}$ with unpolarized electron beams, 
respectively. The solid lines are for $A^b_{1}$ and $A^l_{1}$ while the 
long-dashed lines are for $T^b_{33}$ and $T^l_{33}$.
Two figures present us with several interesting results:
\begin{itemize}
\item The allowed regions strongly depend on electron polarization.
      Combining the bounds obtained with left-handed and right-handed electron
      beams, we obtain very tightly constrained 1-$\sigma$ regions for 
      ${\rm Re}(c_\gamma)$ and ${\rm Re}(c_{_Z})$.
\item Even with unpolarized electrons and positrons, it is possible to obtain
      a closed region for the $CP$-odd parameters by using two or more 
      $CP$-odd asymmetries. 
      The 1-$\sigma$ regions become very loose for the parameter
      ${\rm Re}(c_\gamma)$, but the 1-$\sigma$ regions for  
      ${\rm Re}(c_{_Z})$ remain rather intact. 
\item With polarized electrons, the tightest bound is obtained through 
      the $CP$-odd vector asymmetry $A^b_1$ in the inclusive top-quark decay 
      mode.
\end{itemize}
Numerically, the 1-$\sigma$ allowed region of  
${\rm Re}(c_\gamma)$ and ${\rm Re}(c_{_Z})$ at the c.m. energy 
$\sqrt{s}=500$ GeV with the total $e^+e^-$ integrated 
luminosity 20 fb$^{-1}$, which is the sum of the integrated luminosities
for left- and right-handed electrons, is 
\begin{eqnarray}
|{\rm Re}(c_\gamma)|\leq 0.12,\qquad 
|{\rm Re}(c_{_Z})|\leq 0.20.
\end{eqnarray}

The shadowed parts in Figure~5 show the 1-$\sigma$ allowed 
regions of the $CP$-odd parameters ${\rm Im}(c_\gamma)$ and ${\rm Im}(c_{_Z})$ 
through the $CP$-odd and $CP\tilde{T}$-odd asymmetries (a) $A^b_{E}$, 
$A^b_2$ and $Q^b_{33}$ and (b) $A^l_{E}$, $A^l_2$ and $Q^l_{33}$ with 
polarized electron beams, respectively.
The solid lines with a positive 
(negative) slope are for $A^b_{E}$ and $A^l_{E}$ with right-handed 
(left-handed) electrons while the long-dashed lines with a positive 
(negative) slope are for $A^b_2$ and $A^l_2$ with right-handed (left-handed) 
electrons. And, the dot-dashed lines with a positive (negative) slope are for
$Q^b_{33}$ and $Q^l_{33}$ with right-handed (left-handed) electrons.
On the other hand, the shadowed parts in Figure~6 show 
the 1-$\sigma$ allowed regions of the parameters ${\rm Im}(c_\gamma)$ 
and ${\rm Im}(c_{_Z})$ through (a) $A^b_{E}$, 
$A^b_2$ and $Q^b_{33}$ and (b) $A^l_{E}$, $A^l_2$ and $Q^l_{33}$ with 
unpolarized electron beams, respectively. The solid lines are for $A^b_{E}$ 
and $A^l_{E}$ while the long-dashed lines are for $A^b_2$ and $A^l_2$. 
And, the dot-dashed lines are for $Q^b_{33}$ and $Q^l_{33}$.
Two figures present us with several interesting results:
\begin{itemize}
\item The allowed regions strongly depend on electron polarization.
      Combining the bounds obtained with left-handed and right-handed electron
      beams, we obtain very tightly constrained 1-$\sigma$ regions for 
      ${\rm Im}(c_\gamma)$ and ${\rm Im}(c_{_Z})$.
\item Even with unpolarized electrons and positrons, it is possible to 
      obtain a bounded region for the $CP$-odd parameters by using two or 
      more $CP$-odd asymmetries. We find that the 1-$\sigma$ regions become 
      very loose for the parameter ${\rm Im}(c_{_Z})$, but the 1-$\sigma$ regions 
      for ${\rm Re}(c_\gamma)$ remain rather intact. 
\item With polarized electrons, the tightest bound is obtained through the
      $CP$-odd energy asymmetry $A^b_E$ in the inclusive top-quark decay mode.
\end{itemize}
Numerically, the 1-$\sigma$ allowed region of the parameters 
${\rm Im}(c_\gamma)$ and ${\rm Im}(c_{_Z})$ with the total $e^+e^-$ integrated 
luminosity 20 fb$^{-1}$ at the c.m. energy $\sqrt{s}=500$ GeV is 
\begin{eqnarray}
|{\rm Im}(c_\gamma)|\leq 0.16,\qquad 
|{\rm Im}(c_{_Z})|\leq 0.27.
\end{eqnarray}
%

\section{Compton backscattered laser light}
\label{sec:PLC}

\pr
Let us describe in a general 
framework how photon polarization can provide us with an efficient 
mechanism\cite{BCK} to probe $CP$ invariance
in the two-photon mode. With purely linearly-polarized 
photon beams, we classify all the distributions according to
their $CP$ and $CP\tilde{T}$ properties. 
Then, we show explicitly how linearly 
polarized photon beams allow us to construct two $CP$-odd and 
$CP\tilde{T}$-even asymmetries which do not require  
detailed information on the momenta and polarizations of the 
final-state particles. 

\subsection{Formalism}

\pr
Generally, a purely polarized photon beam state is a linear combination
of two helicity states and the photon polarization vector  
can be expressed in terms of two angles $\alpha$ and $\phi$
in a given coordinate system as
\begin{eqnarray}
|\alpha,\phi\rangle =-\cos(\alpha) e^{-i\phi}|+\rangle
           +\sin(\alpha) e^{i\phi}|-\rangle,
\label{polarization vector}
\end{eqnarray}
where $0\leq \alpha\leq \pi/2$ and $0\leq \phi\leq 2\pi$.
The photon polarization vector (\ref{polarization vector})
implies that the degrees of circular and linear polarization are  
determined by
\begin{eqnarray}
\xi=\cos(2\alpha),\qquad \eta=\sin(2\alpha),
\end{eqnarray}
respectively, and the direction of maximal linear polarization is denoted
by the azimuthal angle $\phi$ in the given coordinate system. 
Note that $\xi^2+\eta^2=1$ as expected
for a purely polarized photon. For a partially polarized photon
beam it is necessary to rescale $\xi$ and $\eta$ by its degree
of polarization $P$ ($0\leq P\leq 1$) as
\begin{eqnarray}
\xi=P\cos(2\alpha),\qquad \eta=P\sin(2\alpha),
\end{eqnarray}
such that $\xi^2+\eta^2=P^2$.

Let us now consider the two-photon system in the c.m. frame 
where two photon momenta are opposing along the $z$-axis. 
The two-photon state vector is 
\begin{eqnarray}
&& |\alpha_1,\phi_1;\alpha_2,\phi_2\rangle
 =|\alpha_1,\phi_1\rangle|\alpha_2,-\phi_2\rangle\nonumber\\ 
&&\hskip 0.5cm  =\cos(\alpha_1)\cos(\alpha_2)\: e^{-i(\phi_1-\phi_2)}|++\rangle
   -\cos(\alpha_1)\sin(\alpha_2)\: e^{-i(\phi_1+\phi_2)}|+-\rangle
    \nonumber\\
&&\hskip 0.7cm -\sin(\alpha_1)\cos(\alpha_2)\: e^{i(\phi_1+\phi_2)}|-+\rangle
   +\sin(\alpha_1)\sin(\alpha_2)\: e^{i(\phi_1-\phi_2)}|--\rangle,
\label{eq:two-photon_wf}
\end{eqnarray}
and then the transition amplitude from the polarized two-photon state 
to a final state $X$ is simply given by
\begin{eqnarray}
\langle X |M|\alpha_1,\phi_1;\alpha_2,\phi_2\rangle .
\end{eqnarray} 
The azimuthal angles $\phi_1$ and $\phi_2$ are the directions
of maximal linear polarization of the two photons, respectively, in 
a common coordinate system (For instance, see Figure~7.). 
In the process $\gamma\gamma \rightarrow t\bar{t}$, 
the scattering plane is taken to be the $x$-$z$ plane
in the actual calculation of the helicity amplitudes. 
The maximal linear polarization angles are then chosen as follows. 
The angle $\phi_1$ ($\phi_2$) is the azimuthal angle of 
the maximal linear polarization of the photon beam, whose momentum 
is in the positive (negative) $z$ direction, with respect to 
the direction of the $t$ momentum in the process 
$\gamma\gamma \rightarrow t\bar{t}$. 
Note that we have used $|\alpha_2,-\phi_2\rangle$ in 
Eq.~(\ref{eq:two-photon_wf}) for the photon whose momentum is along the 
negative $z$ direction in order to employ a common coordinate system 
for the two-photon system. 

For later convenience we introduce the abbreviation
\begin{eqnarray}
M_{\lambda_1\lambda_2}=\langle X|M|\lambda_1\lambda_2\rangle,
\end{eqnarray}
and two angular variables:
\begin{eqnarray}
\chi=\phi_1-\phi_2,\qquad 
\phi=\phi_1+\phi_2,
\end{eqnarray}
where $-2\pi\leq \chi\leq 2\pi$ and $0\leq \phi\leq 4\pi$ 
for a fixed $\chi$.
It should be noted that (i) the azimuthal angle difference, $\chi$, 
is independent of the final state, while the azimuthal angle sum, 
$\phi$, depends on the scattering plane, and (ii) both angles are 
invariant with respect to the Lorentz boost along the two-photon 
beam direction. 

It is straightforward to obtain the angular dependence of the
$\gamma\gamma\rightarrow X$ cross section on the initial beam 
polarizations in terms of the Stokes parameters $(\xi, \bar{\xi})$ 
for the degrees of circular polarization and $(\eta,\bar{\eta})$ for
those of linear polarization of the two initial photon beams, 
respectively, as 
\begin{eqnarray}
&& \Sigma(\xi,\bar{\xi};\eta,\bar{\eta};\chi,\phi)
\equiv \sum_X|\langle X|M|\xi,\bar{\xi};\eta,\bar{\eta};\chi,\phi\rangle |^2,
\label{eq:distribution}
\end{eqnarray}
where the summation over $X$ is for the polarizations of the final states.
Incidentally, the Stokes parameters are expressed in terms of 
two parameters $\alpha_1$ and $\alpha_2$ by   
\begin{eqnarray}
&&\xi=P\cos(2\alpha_1),\qquad \bar{\xi}=\bar{P}\cos(2\alpha_2),
  \nonumber\\
&&\eta=P\sin(2\alpha_1),\qquad \bar{\eta}=\bar{P}\sin(2\alpha_2),
\end{eqnarray}
where $P$ and $\bar{P}$ ($0\leq P,\bar{P}\leq 1$) are the polarization 
degrees of the two colliding photons.
There exist sixteen independent terms, 
all of which are all measurable in polarized two-photon collisions. 
Purely linearly polarized photon beams allow us to 
determine nine terms among all the sixteen terms, while
purely circularly polarized photon beams allow us to determine
only four terms.  The unpolarized cross section is determined in 
both cases. However, both circular and linear polarizations are 
needed to determine the remaining four terms. 

Even though we obtain more information with both circularly and 
linearly polarized beams, we study mainly the case 
where two photons are linearly polarized but not circularly polarized.
The expression of the angular dependence then greatly simplifies to
\begin{eqnarray}
&&{\cal D}(\eta,\bar{\eta};\chi,\phi)
  =\Sigma_{\rm unpol}
  -\frac{1}{2}[\eta\cos(\phi+\chi)+\bar{\eta}\cos(\phi-\chi)]
   \Re(\Sigma_{02})\nonumber\\  
&&\hskip 0.3cm 
  +\frac{1}{2}[\eta\sin(\phi+\chi)-\bar{\eta}\sin(\phi-\chi)]
   \Im(\Sigma_{02})  
  -\frac{1}{2}[\eta\cos(\phi+\chi)-\bar{\eta}\cos(\phi-\chi)]
   \Re(\Delta_{02})\nonumber\\  
&&\hskip 0.3cm 
  +\frac{1}{2}[\eta\sin(\phi+\chi)+\bar{\eta}\sin(\phi-\chi)]
   \Im(\Delta_{02})  
  +\eta\bar{\eta}\cos(2\phi)\Re(\Sigma_{22})
  +\eta\bar{\eta}\sin(2\phi)\Im(\Sigma_{22})\nonumber\\
&&\hskip 0.3cm
  +\eta\bar{\eta}\cos(2\chi)\Re(\Sigma_{00})
  +\eta\bar{\eta}\sin(2\chi)\Im(\Sigma_{00}),
 \label{eq:linear dist}
\end{eqnarray}
where the invariant functions are defined as
\begin{eqnarray}
&&\Sigma_{\rm unpol}=\frac{1}{4}\sum_X
              \left[|M_{++}|^2+|M_{+-}|^2
                   +|M_{-+}|^2+|M_{--}|^2\right]\nonumber\\
&&\Sigma_{02}=\frac{1}{2}\sum_X\left[M_{++}(M^*_{+-}+M^*_{-+})
             +(M_{+-}+M_{-+})M^*_{--}\right]\nonumber\\
&&\Delta_{02}=\frac{1}{2}\sum_X\left[M_{++}(M^*_{+-}-M^*_{-+})
             -(M_{+-}-M_{-+})M^*_{--}\right]\nonumber\\
&&\Sigma_{22}=\frac{1}{2}\sum_X(M_{+-}M^*_{-+}),\qquad\hskip 0.3cm
  \Sigma_{00}=\frac{1}{2}\sum_X(M_{++}M^*_{--}),
\label{quote}
\end{eqnarray}
with the subscripts, $0$ and $2$, representing the magnitude of the 
sum of two photon helicities of the initial two-photon system.

\subsection{Symmetry properties}

\pr
It is useful to classify the invariant functions according to 
their transformation properties under the discrete symmetries,
$CP$ and $CP\tilde{T}$\cite{HPZH}.
We find that $CP$ invariance leads to the relations
\begin{eqnarray}
\sum_X\left(M_{\lambda_1\lambda_2}
            M^*_{\lambda^\prime_1\lambda^\prime_2}\right)
&=&
\sum_X\left(M_{-\lambda_2,-\lambda_1}
            M^*_{-\lambda^\prime_2,-\lambda^\prime_1}\right),\nonumber\\
{\rm d}\sigma(\phi,\chi;\eta,\bar{\eta})&=&
{\rm d}\sigma(\phi,-\chi;\bar{\eta},\eta),
\end{eqnarray}
and, if there are no absorptive parts in the amplitudes, 
$CP\tilde{T}$ invariance leads to the relations
\begin{eqnarray}
\sum_X\left(M_{\lambda_1\lambda_2}
            M^*_{\lambda^\prime_1\lambda^\prime_2}\right)
&=&
\sum_X\left(M^*_{-\lambda_2,-\lambda_1}
            M_{-\lambda^\prime_2,-\lambda^\prime_1}\right),\nonumber \\
{\rm d}\sigma(\phi,\chi;\eta,\bar{\eta})&=&
{\rm d}\sigma(-\phi,\chi;\bar{\eta},\eta).
\end{eqnarray}

The nine invariant functions in Eq.~(\ref{eq:linear dist}) 
can then be divided into four categories
under $CP$ and $CP\tilde{T}$: even-even, even-odd, odd-even,
and odd-odd terms as in Table~2. 
$CP$-odd coefficients directly measure 
$CP$ violation and $CP\tilde{T}$-odd terms indicate rescattering 
effects (absorptive parts in the scattering amplitudes).
Table~2 shows that there exist three $CP$-odd functions; 
$\Im(\Sigma_{02})$, 
$\Im(\Sigma_{00})$ and $\Re(\Delta_{02})$. Here, $\Re$
and $\Im$ are for real and imaginary parts, respectively.
While the first two terms are $CP\tilde{T}$-even, the last 
term $\Re(\Delta_{02})$ is
$CP\tilde{T}$-odd. Since the $CP\tilde{T}$-odd term 
$\Re(\Delta_{02})$ requires the absorptive part in the amplitude,
it is generally expected to be smaller in magnitude than 
the $CP\tilde{T}$-even terms. We therefore study the two $CP$-odd and 
$CP\tilde{T}$-even distributions;
$\Im(\Sigma_{02})$ and $\Im(\Sigma_{00})$.

We can define two $CP$-odd asymmetries from the two distributions, 
$\Im(\Sigma_{02})$ and $\Im(\Sigma_{00})$. 
First, we note that the $\Sigma_{00}$ term does not depend on the 
azimuthal angle $\phi$ whereas the $\Sigma_{02}$ does. 
In order to improve the observability we may integrate 
the $\Im(\Sigma_{02})$ term over the azimuthal angle $\phi$ 
with an appropriate weight function. 
Without any loss of generality we can take $\eta=\bar{\eta}$. 
Then, the quantity $\Im(\Sigma_{00})$ in Eq.~(\ref{eq:linear dist}) 
can be separated by taking the difference of the distributions 
at $\chi=\pm\pi/4$  and the $\Im(\Sigma_{02})$ by taking 
the difference of the distributions at $\chi=\pm\pi/2$.
As a result we obtain the following two integrated $CP$-odd
asymmetries:
\begin{eqnarray}
\hat{A}_{02}=\left(\frac{2}{\pi}\right)
     \frac{\Im(\Sigma_{02})}{\Sigma_{\rm unpol}},\qquad
\hat{A}_{00}=\frac{\Im(\Sigma_{00})}{\Sigma_{\rm unpol}},
\end{eqnarray}
where the factor $(2/\pi)$ in the $\hat{A}_{02}$ stems from taking 
the average over the azimuthal angle $\phi$ with the weight function
${\rm sign}(\cos\phi)$:
\begin{eqnarray}
&&\hat{A}_{02}=\frac{\int^{4\pi}_0{\rm d}\phi [{\rm sign}(\cos\phi)]
\bigg[\left(\frac{{\rm d}\sigma}{{\rm d}\phi}\right)_{\chi=\frac{\pi}{2}}
    -\left(\frac{{\rm d}\sigma}{{\rm d}\phi}\right)_{\chi=-\frac{\pi}{2}}
\bigg]}{\int^{4\pi}_0{\rm d}\phi 
\bigg[\left(\frac{{\rm d}\sigma}{{\rm d}\phi}\right)_{\chi=\frac{\pi}{2}}
    +\left(\frac{{\rm d}\sigma}{{\rm d}\phi}\right)_{\chi=-\frac{\pi}{2}}
  \bigg]},\\
&&\hat{A}_{00}=\frac{\int^{4\pi}_0{\rm d}\phi
\bigg[\left(\frac{{\rm d}\sigma}{{\rm d}\phi}\right)_{\chi=\frac{\pi}{4}}
    -\left(\frac{{\rm d}\sigma}{{\rm d}\phi}\right)_{\chi=-\frac{\pi}{4}}
\bigg]}{\int^{4\pi}_0{\rm d}\phi 
\bigg[\left(\frac{{\rm d}\sigma}{{\rm d}\phi}\right)_{\chi=\frac{\pi}{4}}
    +\left(\frac{{\rm d}\sigma}{{\rm d}\phi}\right)_{\chi=-\frac{\pi}{4}}
\bigg]}.
\end{eqnarray}
In pair production processes such as 
$\gamma\gamma\rightarrow t\bar{t}$, all the distributions, $\Sigma_i$,
can be integrated over the scattering angle $\theta$ with a $CP$-even 
angular cut so as to test $CP$ violation.

\subsection{Photon spectrum}

\pr
Recently, the well-known Compton backscattering\cite{GKS} has drawn a lot of
interest because it can be utilized as a powerful high-energy
photon source at NLC experiments. In this section we give a detailed 
description of the energy spectrum and polarization
of the Compton backscattered laser lights off high energy electrons 
or positrons.

We are interested in the situation where a purely linearly polarized 
laser beam of frequency $\omega_0$ is focused upon an unpolarized electron or 
positron beam of energy $E$. 
In the collision of a laser photon and a linac electron, 
a high energy photon of energy $\omega$, which is partially 
linearly polarized, is emitted at a very small angle, along with the 
scattered electron of energy $E^\prime =E-\omega$. 
The kinematics of the Compton backscattering process is then 
characterized by the dimensionless parameters $x$ and $y$:
\begin{eqnarray}
x=\frac{4E\omega_0}{m^2_e}
 \approx 15.3\left(\frac{E}{\rm TeV}\right)
             \left(\frac{\omega_0}{\rm eV}\right),\qquad
y=\frac{\omega}{E}.
\end{eqnarray}
In general, the backscattered photon energies increase with $x$;
the maximum photon energy fraction is given by
$y_m=x/(1+x)$. Operation below the threshold\cite{GKS} for $e^+e^-$ 
pair production in collisions between the laser beam and the 
Compton-backscattered photon beam requires $x\leq 2(1+\sqrt{2})\approx
4.83$; the lower bound on $x$ depends on the lowest available
laser frequency and the production threshold of a given final
state.

Figure~8(a) shows the photon energy spectrum
for various values of $x$. Clearly large $x$ values are
favored to produce highly energetic photons.
On the other hand, the degree $\eta(y)$ of linear polarization of the 
backscattered photon beam reaches the maximum value at $y=y_m$ 
(See Figure~8(b)),
\begin{eqnarray}
\eta_{\rm max}=\eta(y_m)=\frac{2(1+x)}{1+(1+x)^2},
\end{eqnarray}
and approaches unity for small values of $x$. 
In order to retain large linear polarization  we should keep
the $x$ value as small as possible.

\subsection{Linear polarization transfers}

\pr
In the two-photon collision case only part of linear polarization 
of each incident laser beam is transferred to the high-energy photon beam.
We introduce  two functions, ${\cal A}_\eta$ and ${\cal A}_{\eta\eta}$, 
to denote the degrees of linear polarization transfer\cite{Comment2} 
as
\begin{eqnarray}
{\cal A}_\eta(\tau)=\frac{\langle \phi_0\phi_3\rangle_\tau}{\langle 
               \phi_0\phi_0\rangle_\tau},\qquad 
{\cal A}_{\eta\eta}(\tau)
     =\frac{\langle \phi_3\phi_3\rangle_\tau}{\langle 
               \phi_0\phi_0\rangle_\tau},
\end{eqnarray}
where $\phi_0(y)$ is the photon energy spectrum function and
$\phi_3(y)=2y^2/(x(1-y))^2$ and $\tau$ is the ratio of the $\gamma\gamma$ 
c.m. energy squared $\hat{s}$ to the $e^+e^-$ collider energy squared 
$s$. The function ${\cal A}_\eta$ is for the collision of an 
unpolarized photon beam and a linearly polarized photon beam, and 
the function ${\cal A}_{\eta\eta}$ for the collision of two linearly 
polarized photon beams. The convolution integrals 
$\langle \phi_i\phi_j\rangle_\tau$ ($i,j=0,3$) 
for a fixed value of $\tau$ are defined as  
\begin{eqnarray}
\langle \phi_i\phi_j\rangle_\tau
 =
 \frac{1}{{\cal N}^2(x)}\int^{y_m}_{\tau/y_m}\frac{{\rm d}y}{y}
 \phi_i(y)\phi_j(\tau/y),
\end{eqnarray}
where the normalization factor ${\cal N}(x)$ is by the integral 
of the photon energy spectrum $\phi_0$ over $y$.

The event rates of the $\gamma\gamma\rightarrow X$ reaction with 
polarized photons can be obtained by folding a photon luminosity 
spectral function with the $\gamma\gamma\rightarrow X$ production 
cross section as (for $\eta=\bar{\eta}$) 
\begin{eqnarray}
{\rm d}N_{\gamma\gamma\rightarrow X}
 ={\rm d}L_{\gamma\gamma} 
  {\rm d}\hat{\sigma}(\gamma\gamma\rightarrow X),
\end{eqnarray}
where
\begin{eqnarray}
&& {\rm d}L_{\gamma\gamma}
  =\kappa^2L_{ee}\langle\phi_0\phi_0\rangle_\tau{\rm d}\tau,\\
&& {\rm d}\hat{\sigma}(\gamma\gamma\rightarrow X)=  
   \frac{1}{2\hat{s}}{\rm d}\Phi_X
  \Bigg[\Sigma_{\rm unpol}-\eta{\cal A}_\eta\cos\phi 
        \Re\bigg({\rm e}^{-i\chi}\Sigma_{02}\bigg)\nonumber\\
&& \hskip 0.5cm
 +\eta{\cal A}_\eta\sin\phi \Im\bigg({\rm e}^{-i\chi}\Delta_{02}\bigg)
 +\eta^2{\cal A}_{\eta\eta} 
  \Re\bigg({\rm e}^{-2i\phi}\Sigma_{22}
         +{\rm e}^{-2i\chi}\Sigma_{00}\bigg)\Bigg].
\label{eq:folded dist}
\end{eqnarray}
Here, $\kappa$ is the $e$-$\gamma$ conversion coefficient in
the Compton backscattering and  ${\rm d}\Phi_X$ is the phase space
factor of the final state.
The distribution (\ref{eq:folded dist}) of event rates enables us to 
construct two $CP$-odd asymmetries;
\begin{eqnarray}
A_{02}=\left(\frac{2}{\pi}\right)\frac{N_{02}}{N_{\rm unpol}},\qquad
A_{00}=\frac{N_{00}}{N_{\rm unpol}},
\end{eqnarray}
where with $\tau_{\rm max}=y_m^2$ and $\tau_{\rm min}=M^2_X/s$
we have for the event distributions 
\begin{eqnarray}
  \left(\begin{array}{c}
      N_{\rm unpol} \\
      N_{02}\\
      N_{00}\end{array}\right)
    =\kappa^2L_{ee}\frac{1}{2s}
     \int^{\tau_{\rm max}}_{\tau_{\rm min}}\frac{{\rm d}\tau}{\tau}
     \int {\rm d}\Phi_X 
     \langle\phi_0\phi_0\rangle_\tau
   \left(\begin{array}{c}
     \Sigma_{\rm unpol}\\
     \eta{\cal A}_\eta\Im\left(\Sigma_{02}\right)\\
     \eta^2{\cal A}_{\eta\eta}\Im\left(\Sigma_{00}\right)
       \end{array}\right).
\end{eqnarray}
The asymmetries depend crucially on the two-photon spectrum
and the two linear polarization transfers. 

We first investigate the $\sqrt{\tau}$ dependence of the two-photon 
spectrum and the two linear polarization transfers,
$A_\eta$ and $A_{\eta\eta}$ by varying the value of the dimensionless 
parameter $x$. 
Three values of $x$ are chosen; $x=0.5$, $1$, and $4.83$.
Two figures in Figure~9 clearly show that the energy of two 
photons 
reaches higher ends for larger $x$ values but the maximum linear 
polarization transfers are larger for smaller $x$ values.
We also note that $A_\eta$ (solid lines) is larger than $A_{\eta\eta}$ 
(dashed lines) in the whole range of $\sqrt{\tau}$. 
We should keep the parameter $x$ as large as possible to
reach higher energies. However, larger $CP$-odd asymmetries can be
obtained for smaller $x$ values. Therefore, there should
exist a compromised value of $x$, i.e. the incident laser beam 
frequency $\omega_0$ for the optimal observability of 
$CP$ violation. The energy dependence of the subprocess cross section
and that of the $CP$-odd asymmetries are both essential to find
the optimal $x$ value.

\section{Two-photon mode}
\label{sec:Top_PP}

In this section we reinvestigate $CP$ violation due to the top-quark
EDM in the two-photon mode by extending the previous work\cite{ScKh}
and revising its numerical errors.

\subsection{Helicity Amplitudes}

\pr
The process $\gamma\gamma\rightarrow t\bar{t}$ consists of two Feynman
diagrams and its helicity amplitudes in the $\gamma\gamma$ 
c.m. frame are given by
\begin{eqnarray}
{\cal M}_{\lambda_1\lambda_2;\sigma\bar{\sigma}}
        =
\frac{4\pi\alpha Q^2_tN_c}{(1-\hat{\beta}^2\cos^2\Theta)}
     \left[A_{\lambda_1\lambda_2;\sigma\bar{\sigma}}
         +(i\delta_t)B_{\lambda_1\lambda_2;\sigma\bar{\sigma}}
         +(i\delta_t)^2C_{\lambda_1\lambda_2;\sigma\bar{\sigma}}\right],
\end{eqnarray}
where $\Theta$ is the scattering angle between $t$ and a photon, and
the top-quark EDM factor $\delta_t$ is given by
\begin{eqnarray}
\delta_t=\frac{3}{4}\frac{c_\gamma}{m_t}=\frac{3}{4}d_t^\gamma.
\end{eqnarray}
The SM contributions $A_{\lambda_1\lambda_2;\sigma\bar{\sigma}}$ are given by
\begin{eqnarray}
&&A_{\lambda\lambda;\sigma\sigma}=-\frac{4m_t}{\sqrt{\hat{s}}}
  (\lambda+\sigma\hat{\beta}),\qquad
  A_{\lambda\lambda;\sigma,-\sigma}=0,\nonumber\\
&&A_{\lambda,-\lambda;\sigma\sigma}=\frac{4m_t\hat{\beta}}{\sqrt{\hat{s}}}
  \sigma\sin^2\Theta,\qquad
  A_{\lambda,-\lambda;\sigma,-\sigma}=2\hat{\beta}
  \left(\lambda\sigma+\cos\Theta\right)\sin\Theta,
\end{eqnarray}
the terms $B_{\lambda_1\lambda_2;\sigma\bar{\sigma}}$, which are linear 
in $\delta_t$, and the terms $C_{\lambda_1\lambda_2;\sigma\bar{\sigma}}$,
which are quadratic in $\delta_t$, are given by
\begin{eqnarray}
&&B_{\lambda\lambda;\sigma\sigma}=2\sqrt{\hat{s}}
  \left[\frac{8m^2_t}{\hat{s}}
    +\hat{\beta}(\hat{\beta}-\sigma\lambda)\sin^2\Theta\right],\nonumber\\
&&B_{\lambda\lambda;\sigma,-\sigma}=-4m_t\lambda\hat{\beta}
    \sin\Theta\cos\Theta,\nonumber\\
&&B_{\lambda,-\lambda;\sigma\sigma}=2\sqrt{\hat{s}}\hat{\beta}^2
    \sin^2\Theta,\nonumber\\
&&B_{\lambda,-\lambda;\sigma,-\sigma}=0,
\end{eqnarray}
and
\begin{eqnarray}
&&C_{\lambda\lambda;\sigma\sigma}=-2m_t\sqrt{\hat{s}}\lambda
  \left[\frac{4m^2_t}{\hat{s}}
    +\hat{\beta}(\hat{\beta}-\sigma\lambda)\sin^2\Theta\right],\nonumber\\
&&C_{\lambda\lambda;\sigma,-\sigma}=4m^2_t\hat{\beta}
    \sin\Theta\cos\Theta,\nonumber\\
&&C_{\lambda,-\lambda;\sigma\sigma}=-2m_t\sqrt{\hat{s}}\sigma\hat{\beta}
    \sin^2\Theta,\nonumber\\
&&C_{\lambda,-\lambda;\sigma,-\sigma}=-\hat{s}\hat{\beta}\sin\Theta
  \left[\frac{4m^2_t}{\hat{s}}\cos\Theta
    +\lambda\sigma(1-\hat{\beta}^2\cos^2\Theta)\right],
\end{eqnarray}
where $\lambda,\bar{\lambda}$ and $\sigma/2,\bar{\sigma}/2$
are the two-photon and $t,\bar{t}$ helicities, respectively,
$\hat{s}$ is the $\gamma\gamma$ c.m. 
energy squared, and $\hat{\beta}=\sqrt{1-4m^2_t/\hat{s}}$.

\subsection{Differential cross section}

\pr
In counting experiments where the final $t$ polarizations are not
analyzed, we measure only the following combinations:
\begin{eqnarray}
\sum_X M_{\lambda_1\lambda_2}M^*_{\lambda^\prime_1\lambda^\prime_2}
 =
(eQ_t)^4\sum_{\sigma}\sum_{\bar{\sigma}}
   \tilde{\cal M}_{\lambda_1\lambda_2;\sigma\bar{\sigma}}
   \tilde{\cal M}^*_{\lambda^\prime_1\lambda^\prime_2;\sigma\bar{\sigma}}.
\label{eq:distr6}
\end{eqnarray}
We then find $\Sigma_{\rm unpol}$, $ \Sigma_{02}$, $\Delta_{02}$,
$\Sigma_{22}$, and  $\Sigma_{00}$ from Equation~(\ref{quote}).
The differential cross section for a fixed angle $\chi$ is 
\begin{eqnarray}
&&\!\!\!\frac{{\rm d}^2\sigma}{{\rm d}\cos\Theta{\rm d}\phi}(\chi)
=\frac{\alpha^2Q^4_tN_c\hat{\beta}}{8\hat{s}(1-\hat{\beta}^2\cos^2\Theta)^2}
 \Bigg\{\hat{\Sigma}_{\rm unpol}-\frac{1}{2}\Re\bigg[
 \left(\eta{\rm e}^{-i(\chi+\phi)}
 +\bar{\eta}{\rm e}^{-i(\chi-\phi)}\right)
 \hat{\Sigma}_{02}\bigg]\nonumber\\
&&+\frac{1}{2}\Re\bigg[
  \left(\eta{\rm e}^{-i(\chi+\phi)}
  -\bar{\eta}{\rm e}^{-i(\chi-\phi)}\right)
  \hat{\Delta}_{02}\bigg]+\eta\bar{\eta}\Re\bigg[
  {\rm e}^{-2i\phi}\hat{\Sigma}_{22}+{\rm e}^{-2i\chi}\hat{\Sigma}_{00}
  \bigg]\Bigg\},\\
&&\!\!\!\Sigma_i=\frac{e^4Q^4_t\hat{\Sigma}_i}{(1-\hat{\beta}^2\cos^2\Theta)^2},
   \qquad
\Delta_{02}=\frac{e^4Q^4_t\hat{\Delta}_{02}}{(1-\hat{\beta}^2\cos^2\Theta)^2},
\end{eqnarray}
for $i={\rm unpol},{02},{22}$, and ${00}$.

We first note that all the real parts of the distributions (\ref{eq:distr6}) 
are independent of the anomalous $CP$-odd form factors $c_\gamma$ 
up to linear order
\begin{eqnarray}
&&\hat{\Sigma}_{\rm unpol}=4\left[1+2\hat{\beta}^2\sin^2\Theta
                  -\hat{\beta}^4(1+\sin^4\Theta)\right],\nonumber\\
&&\begin{array}{ll}
  \Re(\hat{\Sigma}_{02})
         =\-\frac{16}{\hat{r}}\hat{\beta}^2\sin^2\Theta,\ \ &
  \Re(\hat{\Delta}_{02})=0,\\
  \Re(\hat{\Sigma}_{22})=-4\hat{\beta}^4\sin^4\Theta,\ \ &
  \Re(\hat{\Sigma}_{00})=-\frac{4}{\hat{r}^2}.
  \end{array} 
\end{eqnarray}
Two $CP$-odd distributions $\Im(\hat{\Sigma}_{02})$ and 
$\Im(\hat{\Sigma}_{00})$ 
have contributions from the $CP$-odd form factor $c_\gamma$
and they are given by
\begin{eqnarray}
\Im(\hat{\Sigma}_{02})=0,\qquad
\Im(\hat{\Sigma}_{00})=\frac{24}{m_t}(1-\beta^2\cos^2\Theta){\rm Re}(c_\gamma).
\label{eq:Main2}
\end{eqnarray}
A few comments on the $CP$-odd distributions are in order.
\begin{itemize}
\item $\Im(\hat{\Sigma}_{02})$ is zero so that it can not be used
      to probe $CP$-violating effects from the real part of the top
      quark EDM.
\item $\Im(\hat{\Sigma}_{00})$ is not suppressed at threshold.
\item The $CP$-odd distribution $\Im(\hat{\Sigma}_{00})$ has the angular 
      dependence ($1-\hat{\beta}^2\cos^2\Theta$) which becomes largest
      at the scattering angle $\Theta=\pi/2$, where the SM contribution 
      is generally small.  We, therefore, expect a large $CP$-odd 
      asymmetry at $\Theta\approx\pi/2$.
\end{itemize}
%

\subsection{Observable consequences of the top-quark EDM}
\label{subsec:observable;top}

\pr
The $CP$-odd distribution $\Im(\Sigma_{02})$ is useless
in determining the top EDM parameter ${\rm Re}(c_\gamma)$. 
because it vanishes in the top-pair production via two-photon fusion
as shown in Eq.~(\ref{eq:Main2}). 
In case of $\Im(\Sigma_{00})$, no spin analysis for the decaying 
top quarks is required and furthermore even the scattering
plane does not need to be identified. 
Even if one excludes the $\tau^+\tau^-+
\not\!{p}$ modes of 1\%, the remaining 99\% of the events can be used
to measure $\Im(\Sigma_{00})$.

We present our numerical results for the experimental parameters
\begin{eqnarray}
\sqrt{s}=0.5\ \ {\rm and}\ \ 1.0\ \ {\rm TeV},\qquad
\kappa^2 L_{ee}=20\ \ {\rm fb}^{-1}.
\end{eqnarray}
The dimensionless parameter $x$, which depends on the laser 
frequency $\omega_0$, is treated as an adjustable parameter.
We note that $\kappa=1$ is the maximally allowed value for the
$e$-$\gamma$ conversion coefficient $\kappa$ and it may be as small
as $\kappa=0.1$ if the collider is optimized for the $e^+e^-$ 
model\cite{GKS}. All one should note is that the significance
of the signal scales as $(\epsilon\cdot\kappa^2\cdot L_{ee})$,
where $\epsilon$ denotes the overall detection efficiency that
is different for $A_{00}$ and $A_{02}$.

The $CP$-odd integrated asymmetry $A_{00}$ depends linearly 
on the form factor ${\rm Re}(c_\gamma)$ in the approximation 
that only the terms linear in the form factor are retained.  
We present the sensitivities to the form factor by varying the
parameter $x$, i.e. the incident laser beam frequency $\omega_0$.

Folding the photon luminosity spectrum and integrating the 
distributions over the polar angle $\Theta$, we obtain the 
$x$-dependence of available event rates:
\begin{eqnarray}
&& \left(\begin{array}{c}
      N_{\rm unpol} \\
      N_{00}\end{array}\right)
    =\kappa^2L_{ee}\frac{\pi\alpha^2Q^4_tN_C}{2s}
     \int^{\tau_{\rm max}}_{\tau_{\rm min}}\frac{{\rm d}\tau}{\tau}
     \int^1_{-1} 
     \frac{\hat{\beta}\langle\phi_0\phi_0\rangle_\tau{\rm d}\cos\Theta}
          {(1-\hat{\beta}^2\cos^2\Theta)^2}
   \left(\begin{array}{c}
     \hat{\Sigma}_{\rm unpol}\\
     {\cal A}_{\eta\eta}\Im\left(\hat{\Sigma}_{00}\right)
       \end{array}\right),
\end{eqnarray}
where $\tau_{max}=(x/(1+x))^2$ and $\tau_{min}=4m^2_t/s$.
After extracting the top EDM form factor ${\rm Re}(c_\gamma)$ from the 
asymmetry $A_{00}$ as
\begin{eqnarray}
A_{00}={\rm Re}(c_\gamma)\tilde{A}_{00},
\end{eqnarray}
we obtain the $1$-$\sigma$ allowed sensitivity of the form factor
${\rm Re}(c_\gamma)$
\begin{eqnarray}
{\rm Max}(|{\rm Re}(c_\gamma)|)
   =\frac{\sqrt{2}}{|\tilde{A}_{00}\sqrt{\varepsilon N_{\rm unpol}}|}, 
\label{eq:EDM_m}
\end{eqnarray}
if no asymmetry is found. Here, $\varepsilon$ denotes the multiplication
of the branching fraction times the experimental detection efficiency.
The $N_{SD}$-$\sigma$ upper bound is determined simply by multiplying 
${\rm Max}(|{\rm Re}(c_\gamma)|)$ by $N_{SD}$.

A crucial issue is to find an  optimal means for maximizing 
the denominator in Eq.~(\ref{eq:EDM_m}) experimentally. 
It requires obtaining the smallest 
possible value of $x$ to make the linear polarization transfer as
large as possible. However, the large top-quark mass does not allow
$x$ to be very small. For a given c.m. energy squared, $s$, 
the allowed range of $x$ is given by
\begin{eqnarray}
\frac{2m_t}{\sqrt{s}-2m_t}\leq x \leq 2(1+\sqrt{2}).
\end{eqnarray}

Experimentally, the process $\gamma\gamma\rightarrow W^+W^-$ is the most severe 
background process against the process $\gamma\gamma\rightarrow t\bar{t}$.
Without a detailed background estimation, we simply take the detection
efficiency $\varepsilon$ to be
\begin{eqnarray}
\varepsilon= 10\%, 
\end{eqnarray}
even though more experimental analyses are required to estimate the efficiency 
precisely. It would be, however, rather straightforward to include the effects 
from any experimental cuts and efficiencies in addition to
the branching factors discussed above. 

Figure~10 shows a very strong $x$ dependence of the 
${\rm Re}(c_\gamma)$ upper bound, ${\rm Max}(|{\rm Re}(c_\gamma)|)$,
at $\sqrt{s}=0.5$ and 1 TeV, from the asymmetry $A_{00}$. The solid line
is for $\sqrt{s}=0.5$ TeV and the long-dashed line for $\sqrt{s}=1$ TeV.
The doubling of $e^+e^-$ c.m. energy improves the sensitivity so much 
and renders the optimal $x$ value smaller than that at $\sqrt{s}=0.5$ TeV.
The $x$ values for the optimal sensitivities 
and the optimal 1-$\sigma$ sensitivitito the $CP$-odd parameter
${\rm Re}(c_\gamma)$ for $\sqrt{s}=0.5$ and 1 TeV are listed in Table~3.

\section{Conclusions}
\label{sec:Top_discussion}

\pr
Large top-quark mass implies that a top quark can serve as an
excellent tool to probe $CP$ violation from new interactions at 
NLC.
 
In the production process $e^+e^-\rightarrow t\bar{t}$, followed
by the $t$ and $\bar{t}$ decays, $CP$ violation from the $T$-odd
top-quark EDM and WDM can be investigated through the angular 
correlations of the $t$ and $\bar{t}$ decay products.

We have completely defined all the available $CP$-odd correlations
and have established the relations between a lot of previously suggested
$CP$-odd correlations and the linearly-independent $CP$-odd correlations.
We have fully analyzed the dependence of all the $CP$-odd observables
on the electron beam polarization.

Most $CP$-odd asymmetries in the process $e^+e^-\rightarrow t\bar{t}$ 
depend on both the top EDM and the top WDM. 
Therefore, the separation of two contributions requires introducing 
electron beam polarization and/or using at least two independent
$CP$-odd observables. 
We found that electron polarization is quite effective in separating 
the top-quark EDM and WDM effects.

In the polarized $\gamma\gamma$ mode, initial $CP$-odd two-photon
polarization configurations allow us to measure the top-quark EDM 
by counting $t\bar{t}$ pair
production events in a straightforward way.
Without any direct information on the momenta of the top-quark
decay products linearly-polarized laser beams with an adjustable 
beam energy provide us with a very efficient way of probing the 
top-quark EDM at a PLC.

The strongest 1-$\sigma$ sensitivity on the top EDM factor
${\rm Re}(c_\gamma)$ for $\sqrt{s}=500$ GeV in the polarized $e^+e^-$ mode 
is obtained through the vector asymmetry $A^b_1$ and, numerically, 
it is $|{\rm Re}(c_\gamma)|\leq 0.13$ for the total $e^+e^-$ 
integrated luminosity 20 fb$^{-1}$.
On the other hand, the optimal 1-$\sigma$ sensitivity on  
${\rm Re}(c_\gamma)$ through the asymmetry $A_{00}$
for $\sqrt{s}=0.5$ TeV in the polarized two-photon mode 
is $|{\rm Re}(c_\gamma)|\leq 0.16$.
Consequently, the polarized $e^+e^-$ mode and the polarized two-photon
mode are competitive in probing $CP$ violation in the top-quark
pair production processes.
Certainly, for more rigorous comparison, we should take the 
momentum-dependent top EDM and WDM into account. 

Soni and Xu\cite{Soni-Xu} have estimated the top EDM factor 
${\rm Re}(c_\gamma)$ in Higgs-boson-exchange models of $CP$ 
nonconservation to be typically of 
the order of $10^{-3}$-$10^{-4}$, which is still much smaller than the
experimental sensitivities in the processes 
$e^+e^-(\gamma\gamma)\rightarrow t\bar{t}$ for the total integrated luminosity 
20 fb$^{-1}$ and the c.m. energy $\sqrt{s}=500$ GeV.  However,  
as indicated in Table~3 and Figure~10,  the two-photon mode 
is expected to greatly improve the experimental constraints on the $T$-odd 
top-quark EDM by increasing the c.m. energy and by adjusting the laser beam 
frequency.

\section*{Acknowledgments}

The authors would like to thank F.~Cuypers, M.~Drees, K.~Hagiwara, 
S.D.~Rindani, H.S.~Song, and P.~Zerwas for helpful discussions. 
The work was supported in part by the KOSEF-DFG large collaboration
project (Project No. 96-0702-01-01-2) and Center for Theoretical
Physics (CTP).
MSB is a postdoctoral fellow supported by KOSEF and Research University
Fund of College of Science at Yonsei University supported by MOE of
Korea.
The work of SYC was supported in part by KOSEF and Korean Federation of 
Science and Technology Societies through the Brain Pool program
and the work of CSK was supported in part by BSRI Program 
(Project No. BSRI-97-2425).

\appendix

\section{The definition and explicit analytic form of ${\cal P}_{\alpha X}$}
\label{appendix:p functions}

The definition of ${\cal P}_{\alpha X}$ ($\alpha=1$ to 16 and $X=L,R$)
in terms of the helicity amplitudes $M^X_{\lambda\bar{\lambda}}$ 
($X=L,R$ and $\lambda,\bar{\lambda}=\pm$) is as follows
\begin{eqnarray}
&&{\cal P}_{1X}=\frac{1}{4}
  \left[|M^X_{++}|^2+|M^X_{--}|^2+|M^X_{+-}|^2+|M^X_{-+}|^2\right],\nonumber\\
&&{\cal P}_{2X}=\frac{1}{3\sqrt{3}}
  \left[{\rm Re}(M^X_{--}M^{X*}_{++})
       +\frac{1}{4}(|M^X_{++}|^2+|M^X_{--}|^2 
                   -|M^X_{+-}|^2-|M^X_{-+}|^2)\right],\nonumber\\
&&{\cal P}_{3X}=\frac{1}{2\sqrt{6}} 
  {\rm Re}\left[(M^X_{++}+M^X_{--})(M^X_{+-}+M^X_{-+})^* \right],\nonumber\\
&&{\cal P}_{4X}= - \frac{1}{2\sqrt{6}}
  {\rm Re}\left[(M^X_{++}-M^X_{--})(M^X_{+-}-M^X_{-+})^* \right],\nonumber\\
&&{\cal P}_{5X}=\frac{1}{2\sqrt{6}}
  {\rm Im}\left[(M^X_{++}+M^X_{--})(M^X_{+-}-M^X_{-+})^* \right],\nonumber\\
&&{\cal P}_{6X}=- \frac{1}{2\sqrt{6}} 
  {\rm Im}\left[(M^X_{++}-M^X_{--})(M^X_{+-}+M^X_{-+})^* \right],\nonumber\\
&&{\cal P}_{7X}=\frac{1}{2\sqrt{6}}
  \left[|M^X_{++}|^2-|M^X_{--}|^2\right],\nonumber\\
&&{\cal P}_{8X}=\frac{1}{2\sqrt{6}}
  \left[|M^X_{+-}|^2-|M^X_{-+}|^2\right],\nonumber\\
&&{\cal P}_{9X}=\frac{1}{3\sqrt{6}}
  \left[{\rm Re}(M^X_{--}M^{X*}_{++})
       -\frac{1}{2}(|M^X_{++}|^2+|M^X_{--}|^2 
                   -|M^X_{+-}|^2-|M^X_{-+}|^2)\right],\nonumber\\
&&{\cal P}_{10X}=\frac{1}{3\sqrt{2}}
               {\rm Re}\left(M^X_{-+}M^{X*}_{+-}\right),\nonumber\\
&&{\cal P}_{11X}=\frac{1}{3\sqrt{2}}
               {\rm Im}\left(M^X_{-+}M^{X*}_{+-}\right),\nonumber\\
&&{\cal P}_{12X}=\frac{1}{3\sqrt{2}}
               {\rm Im}\left(M^X_{++}M^{X*}_{--}\right),\nonumber\\
&&{\cal P}_{13X}=\frac{1}{6\sqrt{2}}
  {\rm Im}\left[(M^X_{++}-M^X_{--})(M^X_{+-}-M^X_{-+})^* \right],\nonumber\\
&&{\cal P}_{14X}= - \frac{1}{6\sqrt{2}}
  {\rm Im}\left[(M^X_{++}+M^X_{--})(M^X_{+-}+M^X_{-+})^* \right],\nonumber\\
&&{\cal P}_{15X}=\frac{1}{6\sqrt{2}}
  {\rm Re}\left[(M^X_{++}-M^X_{--})(M^X_{+-}+M^X_{-+})^* \right],\nonumber\\
&&{\cal P}_{16X}=\frac{1}{6\sqrt{2}}
  {\rm Re}\left[(M^X_{++}+M^X_{--})(M^X_{+-}-M^X_{-+})^* \right].
\end{eqnarray}

It is simple to  derive 
all the ${\cal P}_{\alpha X}$ terms up to linear in $c_\gamma$ and $c_{_Z}$
from the helicity amplitudes of $e^+e^-\rightarrow t\bar{t}$, 
neglecting higher-order terms in the form factors. 
In the linear approximation, the $CP$-even and $CP\tilde{T}$-even terms
independent of $c_\gamma$ and $c_{_Z}$ are given by
\begin{eqnarray}
{\cal P}_{1L,R}&=&\frac{1}{2}(v^2_{L,R}+\beta^2a^2_{L,R})(1+\cos^2\Theta)
               \mp 2\beta v_{L,R}a_{L,R}\cos\Theta \nonumber \\
   && \hskip 1cm
               +\frac{1}{2}(1-\beta^2)v^2_{L,R}\sin^2\Theta,\nonumber\\
{\cal P}_{2L,R}&=&-\frac{1}{3\sqrt{3}}\left[
               \frac{1}{2}(v^2_{L,R}+\beta^2a^2_{L,R})(1+\cos^2\Theta)
               \mp 2\beta v_{L,R}a_{L,R}\cos\Theta \right. \nonumber \\
  && \hskip 1cm
  \left. +\, \frac{1}{2}(1-\beta^2)v^2_{L,R}\sin^2\Theta\right],\nonumber\\
{\cal P}_{4L,R}&=& - \frac{2}{\sqrt{6}\gamma}v_{L,R}(\beta a_{L,R}\cos\Theta\mp
               v_{L,R}) \sin\Theta,\nonumber\\
{\cal P}_{8L,R}&=& \frac{2}{\sqrt{6}}\left[ \beta v_{L,R}a_{L,R}
              (1 + \cos^2\Theta) \mp (v^2_{L,R} + \beta^2a^2_{L,R})\cos\Theta
              \right],\nonumber\\
{\cal P}_{9L,R}&=&\frac{1}{3\sqrt{6}}
                \left[(v^2_{L,R}+\beta^2a^2_{L,R})(1+\cos^2\Theta)
               \mp 4\beta v_{L,R}a_{L,R}\cos\Theta \right.\nonumber \\
   && \hskip 1cm \left.
               -2(1-\beta^2)v^2_{L,R}\sin^2\Theta\right],\nonumber\\
{\cal P}_{10L,R}&=&-\frac{1}{3\sqrt{2}}
                 (v^2_{L,R}-\beta^2a^2_{L,R})\sin^2\Theta,\nonumber\\
{\cal P}_{15L,R}&=&\frac{\sqrt{2}}{3\gamma}v_{L,R}
                (v_{L,R}\cos\Theta\mp\beta a_{L,R})\sin\Theta,
\end{eqnarray}
where $v_{L,R}$, $a_{L,R}$ and $c_{L,R}$ are defined in 
Eqs.~(\ref{eq:coupling_def1}) and (\ref{eq:coupling_def2}).

Every $CP$-even and $CP\tilde{T}$-odd term vanishes at 
the tree level:
\begin{eqnarray}
  {\cal P}_{6L}={\cal P}_{6R}={\cal P}_{11L}
 ={\cal P}_{11R}={\cal P}_{13L}={\cal P}_{13R}=0.
\end{eqnarray}
These $\tilde{T}$-odd terms can have finite contributions
from QCD or QED loop corrections\cite{KLY} through the absorptive
parts in the amplitude so that the terms can provide an 
important QCD test since the dominant contributions are 
from one-loop QCD contributions.   

Every $CP$-odd and $CP\tilde{T}$-even term 
${\cal P}_{\alpha X}$, which depends on the real parts of  
$c_\gamma$ and $c_{_Z}$ is given by 
\begin{eqnarray}
&&{\cal P}_{5L,R}= 
  \frac{\sqrt{6}}{3}\gamma\beta (\beta a_{L,R}\cos\Theta\mp v_{L,R}) 
      \sin\Theta{\rm Re}(c_{L,R}),\nonumber\\
&&{\cal P}_{12L,R}=
  - \frac{\sqrt{2}}{6}\beta v_{L,R}\sin^2\Theta{\rm Re}(c_{L,R}),\nonumber\\
&&{\cal P}_{14L,R}=
  -\frac{\sqrt{2}}{3}\gamma\beta(v_{L,R}\cos\Theta\mp\beta a_{L,R})
      \sin\Theta{\rm Re}(c_{L,R}),
\end{eqnarray}
whereas every $CP$-odd and $CP\tilde{T}$-odd term 
${\cal P}_{\alpha X}$, which depends on the imaginary parts of the 
form factors, $c_\gamma$ and $c_{_Z}$, is given by 
\begin{eqnarray}
&& {\cal P}_{3L,R} =
  - \frac{\sqrt{6}}{3}\gamma\beta(v_{L,R}\cos\Theta\mp\beta a_{L,R}) 
  \sin\Theta{\rm Im}(c_{L,R}),\nonumber\\
&& {\cal P}_{7L,R} =
  - \frac{\sqrt{6}}{3}\beta v_{L,R}\sin^2\Theta{\rm Im}(c_{L,R}),\nonumber\\
&& {\cal P}_{16L,R}=
  - \frac{\sqrt{2}}{3}\gamma\beta(\beta a_{L,R}\cos\Theta\mp v_{L,R})
  \sin\Theta{\rm Im}(c_{L,R}).
\end{eqnarray}

\section{The definition of angular correlations
${\cal D}_{\alpha}$ and ${\cal D}^\prime_{\beta}$}
\label{appendix:cal_D functions}

For notational convenience we use the following abbreviations
\begin{eqnarray}
&&\xi_1=\sin\theta\cos\phi,\qquad 
  \xi_2=\sin\theta\sin\phi,\qquad 
  \xi_3=\cos\theta,\nonumber\\
&&\bar{\xi}_1=\sin\bar{\theta}\cos\bar{\phi},\qquad   
  \bar{\xi}_2=\sin\bar{\theta}\sin\bar{\phi},\qquad  
  \bar{\xi}_3=\cos\bar{\theta}.
\end{eqnarray}
Then the orthornormal decay angular correlations ${\cal D}_\alpha$ 
($\alpha=1$ to $16$) and ${\cal D}^\prime_\beta$ ($\beta=1$ to $12$)
are defined in terms of $\xi_i$ and $\bar{\xi}_i$ ($i=1,2,3$) as 
\begin{eqnarray}
&&{\cal D}_1=1, \hskip 5.5 cm 
  {\cal D}_2=\sqrt{3}(\xi_1 \bar\xi_1+\xi_2 \bar\xi_2+\xi_3 \bar\xi_3),
    \nonumber\\ 
&&{\cal D}_3=\frac{\sqrt{3}}{\sqrt{2}}(\xi_1+\bar\xi_1), \hskip 3.7cm
  {\cal D}_4=\frac{\sqrt{3}}{\sqrt{2}}(\xi_1-\bar\xi_1), \nonumber\\ 
&&{\cal D}_5=\frac{\sqrt{3}}{\sqrt{2}}(\xi_2+\bar\xi_2),  \hskip 3.7cm
  {\cal D}_6=\frac{\sqrt{3}}{\sqrt{2}}(\xi_2-\bar\xi_2) \nonumber\\ 
&&{\cal D}_7=\frac{\sqrt{3}}{\sqrt{2}}(\xi_3+\bar\xi_3), \hskip 3.7cm
  {\cal D}_8=\frac{\sqrt{3}}{\sqrt{2}}(\xi_3-\bar\xi_3), \nonumber\\ 
&&{\cal D}_9=\frac{3}{\sqrt{6}}(\xi_1 \bar\xi_1+\xi_2 \bar\xi_2 
               -2\xi_3 \bar\xi_3),  \hskip 1.7cm
  {\cal D}_{10}=\frac{3}{\sqrt{2}}(\xi_1 \bar\xi_1-\xi_2 \bar\xi_2),\nonumber\\ 
&&{\cal D}_{11}=\frac{3}{\sqrt{2}}(\xi_1 \bar\xi_2+\xi_2 \bar\xi_1) , 
                  \hskip 2.9 cm
  {\cal D}_{12}=\frac{3}{\sqrt{2}}(\xi_1 \bar\xi_2-\xi_2 \bar\xi_1), \nonumber\\ 
&&{\cal D}_{13}=\frac{3}{\sqrt{2}}(\xi_2 \bar\xi_3+\xi_3 \bar\xi_2), 
                  \hskip 2.9 cm
  {\cal D}_{14}=\frac{3}{\sqrt{2}}(\xi_2 \bar\xi_3-\xi_3 \bar\xi_2), \nonumber\\ 
&&{\cal D}_{15}=\frac{3}{\sqrt{2}}(\xi_3 \bar\xi_1+\xi_1 \bar\xi_3),  
                  \hskip 2.9 cm
  {\cal D}_{16}=\frac{3}{\sqrt{2}}(\xi_3 \bar\xi_1-\xi_1 \bar\xi_3), 
\end{eqnarray}
and
\begin{eqnarray}
&&{\cal D}^\prime_1=\frac{3 \sqrt{5}}{2 \sqrt{2}}(-\xi_3^2+\bar\xi_3^2), 
                     \hskip 3.2cm
  {\cal D}^\prime_2=\frac{\sqrt{15}}{2\sqrt{2}}(\xi_1^2-\xi_2^2-\bar\xi_1^2 
                   +\bar\xi_2^2),\nonumber\\
&&{\cal D}^\prime_3=\frac{\sqrt{15}}{\sqrt{2}}(\xi_1 \xi_2-\bar\xi_1 \bar\xi_2),
                     \hskip 2.9cm
  {\cal D}^\prime_4=\frac{\sqrt{15}}{\sqrt{2}}(\xi_2 \xi_3-\bar\xi_2 \bar\xi_3),
                    \nonumber\\
&&{\cal D}^\prime_5=\frac{\sqrt{15}}{\sqrt{2}}(\xi_3 \xi_1-\bar\xi_3 \bar\xi_1),
                    \hskip 2.9cm
  {\cal D}^\prime_6=\frac{3\sqrt{5}}{2\sqrt{2}}\left[\xi_2(\bar\xi_1^2-\bar\xi_2^2) 
                   +(\xi_1^2-\xi_2^2) \bar\xi_2\right],  \nonumber\\ 
&&{\cal D}^\prime_7=\frac{3 \sqrt{5}}{\sqrt{2}}\xi_1 \bar\xi_1(\xi_2+\bar\xi_2), 
                     \hskip 2.9cm
  {\cal D}^\prime_8=\frac{\sqrt{15}}{2 \sqrt{2}}\left[\xi_2(1-3 \bar\xi_3^2) 
                   +(1-3 \xi_3^2) \bar\xi_2\right],\nonumber\\  
&&{\cal D}^\prime_9=\frac{3 \sqrt{5}}{\sqrt{2}}\xi_3 \bar\xi_3(\xi_2+\bar\xi_2),
                    \hskip 2.9cm
  {\cal D}^\prime_{10}=\frac{3 \sqrt{5}}{\sqrt{2}}(\xi_1 \bar\xi_2 \bar\xi_3 
                     +\bar\xi_1 \xi_2 \xi_3), \nonumber\\
&&{\cal D}^\prime_{11}=\frac{3 \sqrt{5}}{\sqrt{2}}(\xi_2 \bar\xi_3 \bar\xi_1 
                     +\bar\xi_2 \xi_3 \xi_1), \hskip 2.2cm 
  {\cal D}^\prime_{12}=\frac{3 \sqrt{5}}{\sqrt{2}}(\xi_3 \bar\xi_1 \bar\xi_2 
                     +\bar\xi_3 \xi_1 \xi_2),   
\end{eqnarray}
The correlation functions ${\cal D}$ and ${\cal D}^\prime$
are normalized to satisfy the orthonormality conditions;
\begin{eqnarray}
&&\langle {\cal D}_\alpha {\cal D}_{\alpha^\prime}\rangle\equiv
  \frac{1}{(4\pi)^2}\int\int{\rm d}\Omega{\rm d}\bar{\Omega}
        {\cal D}_\alpha {\cal D}_{\alpha^\prime} =
        \delta_{\alpha\alpha^\prime},\nonumber\\
&&\langle {\cal D}_\alpha {\cal D}^\prime_{\beta^\prime}\rangle\equiv
  \frac{1}{(4\pi)^2}\int\int{\rm d}\Omega{\rm d}\bar{\Omega}
        {\cal D}_\alpha {\cal D}^\prime_{\beta^\prime} =
        0,\nonumber\\
&&\langle {\cal D}^\prime_\beta {\cal D}^\prime_{\beta^\prime}\rangle\equiv
  \frac{1}{(4\pi)^2}\int\int{\rm d}\Omega{\rm d}\bar{\Omega}
        {\cal D}^\prime_\beta {\cal D}^\prime_{\beta^\prime} =
        \delta_{\beta\beta^\prime},
\end{eqnarray}
where $\alpha (\alpha^\prime)=1$ to $16$ 
and $\beta (\beta^\prime)=1$ to $12$.

\section*{References}

\newcommand{\prd}[1]{Phys.~Rev.~D{{\bf #1}}}
\newcommand{\prl}[1]{Phys.~Rev.~Lett.~{{\bf #1}}}
\newcommand{\plb}[1]{Phys.~Lett.~B{{\bf #1}}}
\newcommand{\npb}[1]{Nucl.~Phys.~B{{\bf #1}}}
\newcommand{\zpc}[1]{Z.~Phys.~C{{\bf #1}}}
\newcommand{\progtp}[1]{Prog.~Theor.~Phys.~{{\bf #1}}}
\newcommand{\jetpl}[1]{JETP Lett.~{{\bf #1}}}
\newcommand{\sjnp}[1]{Sov.~J.~Nucl.~Phys.~{{\bf #1}}}

\newpage \section*{Tables}

\begin{itemize}

\item[{\bf Table~1}:] $CP$ and $CP\tilde{T}$ properties of 
      ${\cal P}_{\alpha X}$'s and ${\cal D}_\alpha$'s ($X=L,R$ and $\alpha=1$ 
      to $16$).  

\item [{\bf Table~2}:] $CP$ and $CP\tilde{T}$ properties of the invariant 
      functions and the angular distributions.

\item[{\bf Table~3}:] The optimal 1-$\sigma$ sensitivities to the $CP$-odd 
      top EDM form factor ${\rm Re}(c_\gamma)$ and their corresponding $x$
      values for $\sqrt{s}=0.5$ and 1 TeV.

\end{itemize}

\vskip 3cm
\section*{Figures}
\begin{itemize}
\item[{\bf Figure~1}:] Feynman diagram for the $Vtt$ ($V=\gamma, Z$) 
      vertex.

\item[{\bf Figure~2}:] Schematic view of the sequential processes 
      $e^+e^- \rightarrow t\bar{t} \rightarrow (bW^+)(\bar{b}W^-) \rightarrow 
      (bl^+\nu_l)(\bar{b}l^-\bar{\nu}_l)$. The dashed lines are for 
      invisible particle trajectories in a particle detector.

\item[{\bf Figure~3}:] The 1-$\sigma$ allowed region of the $CP$-odd 
      parameters ${\rm Re}(c_\gamma$) and ${\rm Re}(c_{_Z})$ through 
      the $CP$-odd 
      and $CP\tilde{T}$-even asymmetries (a) $A^b_{1}$ and $T^b_{33}$ and 
      (b) $A^l_{1}$ and $T^l_{33}$ with polarized electron beams, respectively, 
      for the $e^+e^-$ integrated luminosity 10 fb$^{-1}$ and for the c.m. 
      energy $\sqrt{s}=500$ GeV. 
      The solid lines with a positive (negative) slope are for $A^b_{1}$
      and $A^l_{1}$ with right-handed (left-handed) electrons while the 
      long-dashed lines with a positive (negative) slope are for $T^b_{33}$ 
      and $T^l_{33}$ with right-handed (left-handed) electrons.

\item[{\bf Figure~4}:] The 1-$\sigma$ allowed region of the $CP$-odd 
      parameters ${\rm Re}(c_\gamma)$ and ${\rm Re}(c_{_Z})$ through 
      the $CP$-odd 
      and $CP\tilde{T}$-even asymmetries (a) $A^b_{1}$ and $T^b_{33}$ and 
      (b) $A^l_{1}$ and $T^l_{33}$ with unpolarized electron beams, 
      respectively, 
      for the $e^+e^-$ integrated luminosity 20 fb$^{-1}$ and for the c.m. 
      energy $\sqrt{s}=500$ GeV. 
      The solid lines are for $A^b_{1}$ and $A^l_{1}$ while the long-dashed 
      lines are for $T^b_{33}$ and $T^l_{33}$. 

\item[{\bf Figure~5}:] The 1-$\sigma$ allowed region of the $CP$-odd 
      parameters ${\rm Im}(c_\gamma)$ and ${\rm Im}(c_{_Z})$ through 
      the $CP$-odd 
      and $CP\tilde{T}$-odd asymmetries (a) $A^b_{E}$, $A^b_2$ and $Q^b_{33}$ 
      and (b) $A^l_{E}$, $A^l_2$ and $Q^l_{33}$ with polarized electron beams, 
      respectively, for the $e^+e^-$ integrated luminosity 10 fb$^{-1}$ and 
      for the c.m. energy $\sqrt{s}=500 GeV$. 
      The solid lines with a positive (negative) slope are for $A^b_{E}$
      and $A^l_{E}$ with right-handed (left-handed) electrons while the 
      long-dashed lines with a positive (negative) slope are for $A^b_2$ 
      and $A^l_2$ with right-handed (left-handed) electrons. 
      And, the dot-dashed lines with a positive (negative) slope are for
      $Q^b_{33}$ and $Q^l_{33}$ with right-handed (left-handed) electrons.

\item[{\bf Figure~6}:] The 1-$\sigma$ allowed region of the $CP$-odd 
      parameters ${\rm Im}(c_\gamma)$ and ${\rm Im}(c_{_Z})$ through 
      the $CP$-odd 
      and $CP\tilde{T}$-odd asymmetries (a) $A^b_{E}$, $A^b_2$ and $Q^b_{33}$ 
      and (b) $A^l_{E}$, $A^l_2$ and $Q^l_{33}$ with unpolarized electron 
      beams, respectively, for the $e^+e^-$ integrated luminosity 
      20 fb$^{-1}$ and for the c.m. energy $\sqrt{s}=500$ GeV. 
      The solid lines are for $A^b_{E}$ 
      and $A^l_{E}$ while the long-dashed lines are for $A^b_2$ and $A^l_2$. 
      And, the dashed lines are for $Q^b_{33}$ and $Q^l_{33}$.

\item[{\bf Figure~7}:] The coordinate system in the colliding 
      $\gamma\gamma$ c.m. frame. The scattering angle, $\Theta$, and 
      the azimuthal angles, $\phi_1$ and $\phi_2$, for the linear
      polarization directions measured from the scattering plane
      are described.

\item[{\bf Figure~8}:] (a) the photon energy spectrum and 
       (b) the degree of linear polarization of the Compton backscattered 
       photon beam for $x=4E\omega_0/m^2_e=0.5$, $1$ and $4.83$.

\item[{\bf Figure~9}:] (a) the $\gamma\gamma$ luminosity spectrum 
      and (b) the two linear polarization transfers, $A_\eta$ (solid lines)
      and $A_{\eta\eta}$ (dashed lines), for $x=4E\omega_0/m^2_e=0.5$, 
      $1$ and $4.83$.

\item[{\bf Figure~10}:] The $x$ dependence of the ${\rm Re}(c_\gamma)$ 
      upper bound, Max($|{\rm Re}(c_\gamma)|$), at $\sqrt{s}=0.5$ and 1 TeV, 
      from the asymmetry $A_{00}$. The solid line is for $\sqrt{s}=0.5$ TeV 
      and the long-dashed line for $\sqrt{s}=1$ TeV.

\end{itemize} 

\newpage
\mbox{ }
\hskip 2.5cm

\begin{center}
{\bf\large Table~1}
\end{center}
\begin{center}
\begin{tabular}{|c|c|c|c|c|}\hline
 $CP$  & $CP\tilde{T}$ & ${\cal P}_{\alpha X}$ 
         &$ {\cal D}_\alpha$ & Number  \\
\hline
even & even & ${\cal P}_{1X},{\cal P}_{2X},{\cal P}_{4X}, {\cal P}_{8X}$
            & ${\cal D}_1,{\cal D}_2,{\cal D}_4, {\cal D}_8$
            &  7  \\ 
{ }  & { }  & ${\cal P}_{9X},{\cal P}_{10X},{\cal P}_{15X}      $ 
            & ${\cal D}_9,{\cal D}_{10},{\cal D}_{15}      $ 
            & { } \\ \hline
even & odd  & ${\cal P}_{6X},{\cal P}_{11X}, {\cal P}_{13X}     $
            & ${\cal D}_6,{\cal D}_{11}, {\cal D}_{13}     $
            & 3  \\ \hline
odd  & even & ${\cal P}_{5X},{\cal P}_{12X}, {\cal P}_{14X}     $
            & ${\cal D}_5,{\cal D}_{12}, {\cal D}_{14}     $
            & 3  \\ \hline
odd  & odd  & ${\cal P}_{3X}, {\cal P}_{7X}, {\cal P}_{16X}     $
            & ${\cal D}_3, {\cal D}_{7}, {\cal D}_{16}     $
            & 3  \\
\hline
\end{tabular}
\end{center}

\vskip 3cm
\begin{center}
{\bf\large Table~2}
\end{center}
\begin{center}
\begin{tabular}{|c|c|c|c|}\hline
 \mbox{ }\hskip 0.2cm $CP$ \mbox{ }\hskip 0.2cm   
&\mbox{ }\hskip 0.2cm $CP\tilde{T}$\mbox{ }\hskip 0.2cm   
&\mbox{ }\hskip 0.2cm Invariant functions \mbox{ }\hskip 0.2cm  
&\mbox{ }\hskip 0.2cm Angular dependences \mbox{ }\hskip 0.2cm  \\
\hline
even & even & $\Sigma_{\rm unpol}$     
            & { } \\
            \cline{3-4}
{ }  & { }  & $\Re(\Sigma_{02})$  
            & $\eta\cos(\phi+\chi)+\bar{\eta}\cos(\phi-\chi)$  \\
            \cline{3-4}
{ }  & { }  & $\Re(\Sigma_{22})$
            & $\eta\bar{\eta}\cos(2\phi)$ \\
            \cline{3-4}
{ }  & { }  & $\Re(\Sigma_{00})$  
            & $\eta\bar{\eta}\cos(2\chi)$ \\ \hline 
even & odd  & $\Im(\Delta_{02})$
            & $\eta\sin(\phi+\chi)+\bar{\eta}\sin(\phi-\chi)$  \\
            \cline{3-4}
{ }  & { }  & $\Im(\Sigma_{22})$  
            & $\eta\bar{\eta}\sin(2\phi)$ \\ \hline
odd  & even & $\Im(\Sigma_{02})$
            & $\eta\sin(\phi+\chi)-\bar{\eta}\sin(\phi-\chi)$  \\
            \cline{3-4}
{ }  & { }  & $\Im(\Sigma_{00})$  
            & $\eta\bar{\eta}\sin(2\chi)$  \\ \hline
odd  & odd  & $\Re(\Delta_{02})$  
            & $\eta\cos(\phi+\chi)-\bar{\eta}\cos(\phi-\chi)$  \\
\hline
\end{tabular}
\end{center}

\vskip 3cm

\begin{center}
{\bf\large Table~3}
\end{center}
\begin{center}
\begin{tabular}{|c|c|c|}\hline
   $\sqrt{s}$        &   0.5   &  1.0  \\ \hline
       $x$           &   3.43  &  0.85 \\ 
${\rm Re}(c_\gamma)$ &   0.16  &  0.02 \\ \hline
\end{tabular}
\end{center}

\newpage  
\mbox{ }
\vskip 3cm

\begin{figure}[h]
\hbox to\textwidth{\hss\epsfig{file=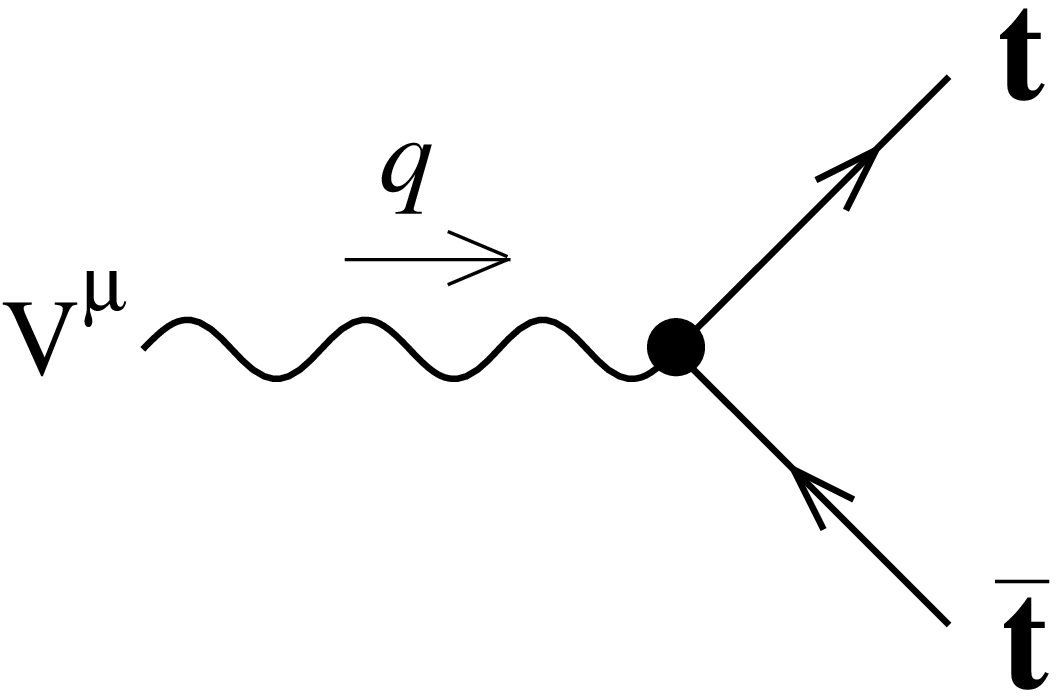,width=15cm,height=10cm}\hss}
\end{figure}

\vskip 3cm
\begin{center}
{\bf\large Figure~1}
\end{center} 

\newpage
\mbox{ }
\vskip 3cm
\begin{figure}[h]
\hbox to\textwidth{\hss\epsfig{file=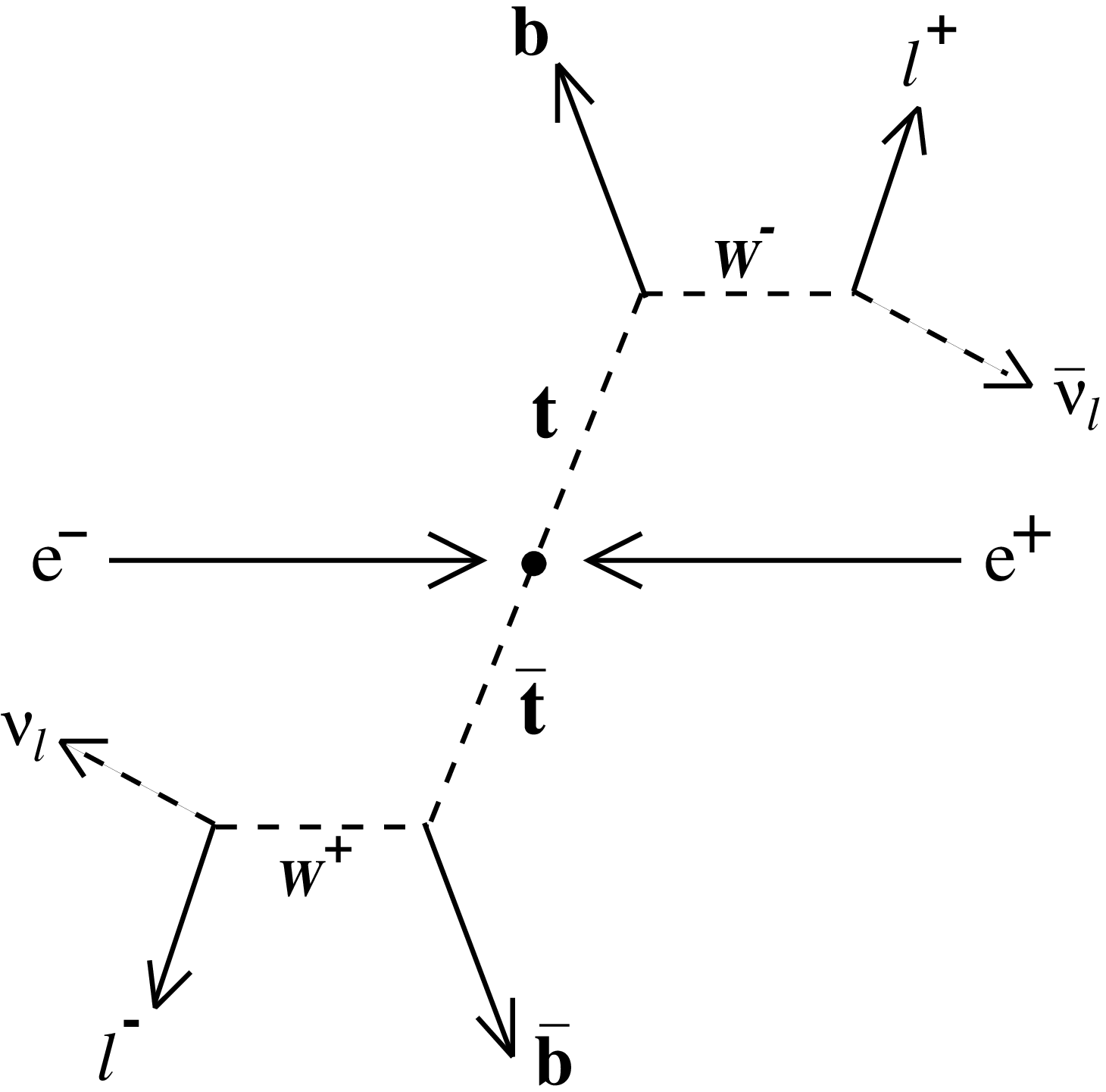,width=13cm,height=11cm}\hss}
\end{figure}

\vskip 3cm
\begin{center}
{\bf\large Figure~2}
\end{center} 

\newpage
\mbox{ }
\vskip 3cm
\begin{figure}[h]
\hbox to\textwidth{\hss\epsfig{file=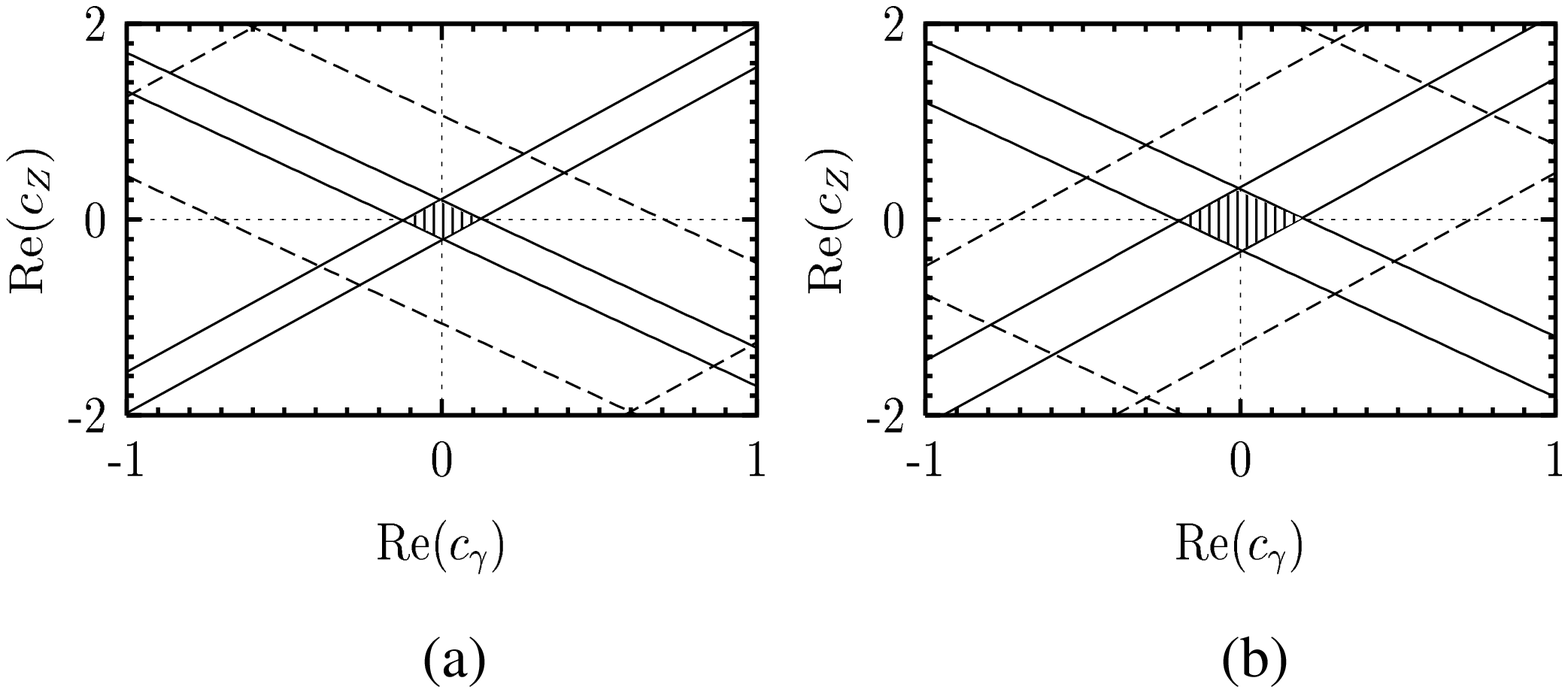,width=15cm,height=10cm}\hss}
\end{figure}

\vskip 3cm
\begin{center}
{\bf\large Figure~3}
\end{center} 

\newpage
\mbox{ }
\vskip 3cm
\begin{figure}[h]
\hbox to\textwidth{\hss\epsfig{file=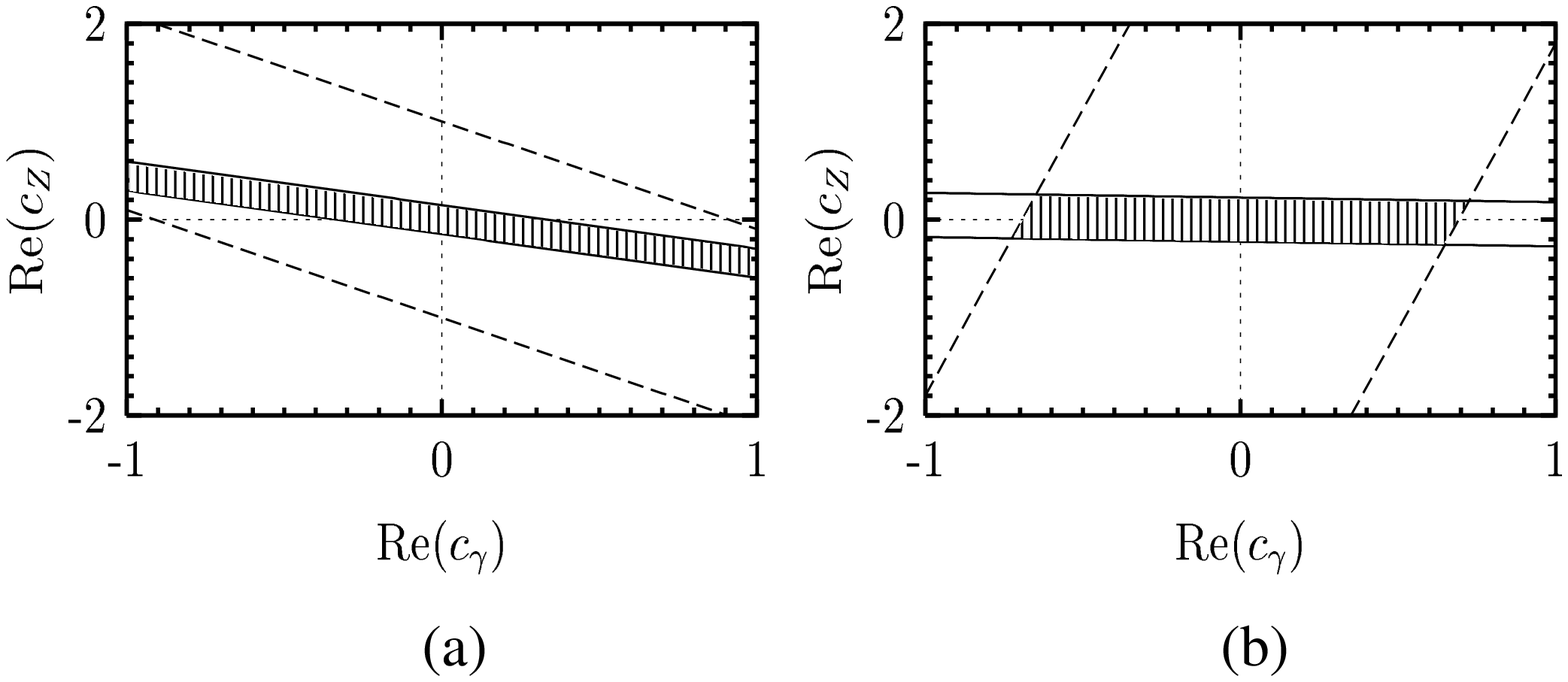,width=15cm,height=10cm}\hss}
\end{figure}

\vskip 3cm
\begin{center}
{\bf\large Figure~4}
\end{center} 

\newpage
\mbox{ }
\vskip 3cm
\begin{figure}[h]
\hbox to\textwidth{\hss\epsfig{file=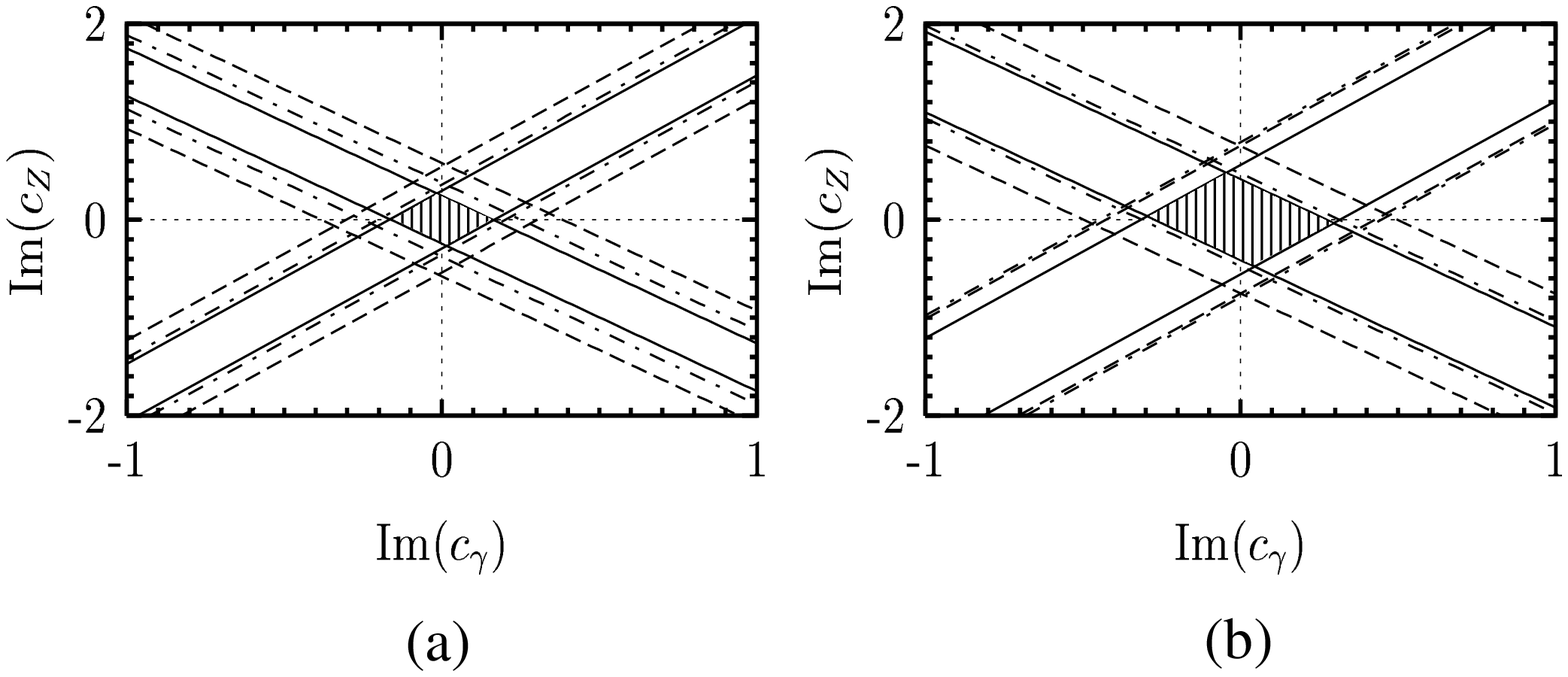,width=15cm,height=10cm}\hss}
\end{figure}

\vskip 3cm
\begin{center}
{\bf\large Figure~5}
\end{center} 

\newpage
\mbox{ }
\vskip 3cm
\begin{figure}[h]
\hbox to\textwidth{\hss\epsfig{file=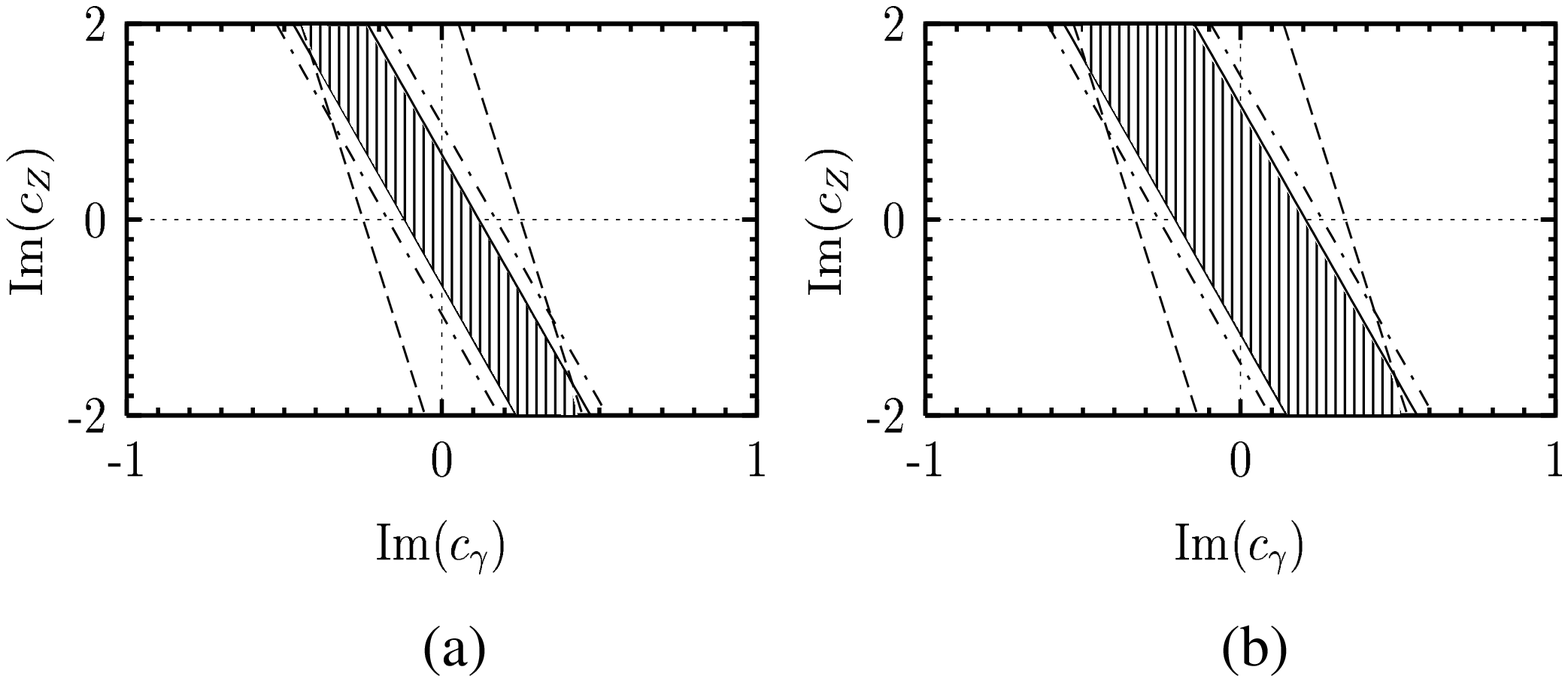,width=15cm,height=10cm}\hss}
\end{figure}

\vskip 3cm
\begin{center}
{\bf\large Figure~6}
\end{center} 

\newpage
\mbox{ }
\vskip 3cm
\begin{figure}[h]
\hbox to\textwidth{\hss\epsfig{file=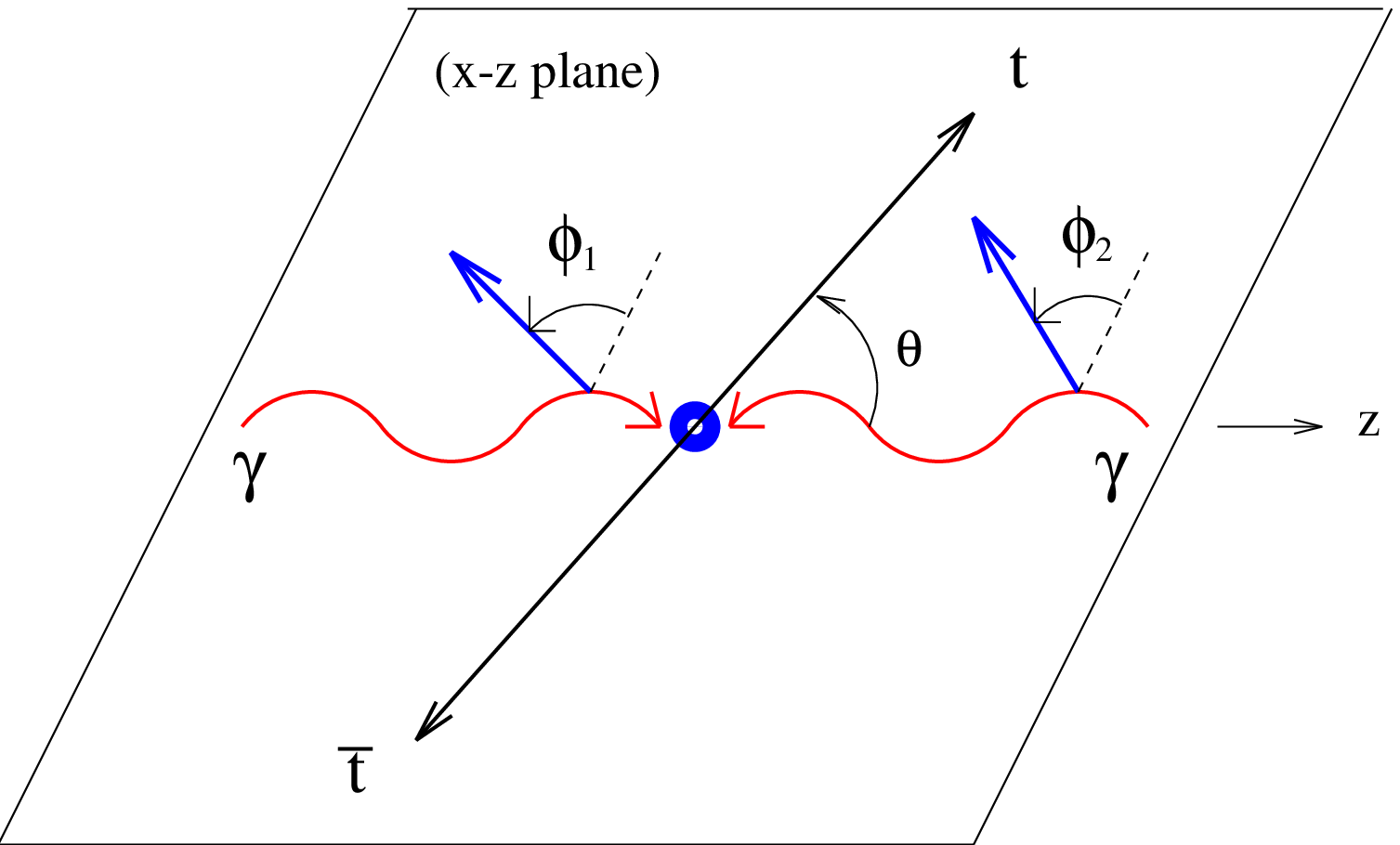,width=15cm,height=11cm}\hss}
\end{figure}

\vskip 3cm
\begin{center}
{\bf\large Figure~7}
\end{center}

\newpage
\mbox{ }
\vskip 3cm
\begin{figure}[h]
\hbox to\textwidth{\hss\epsfig{file=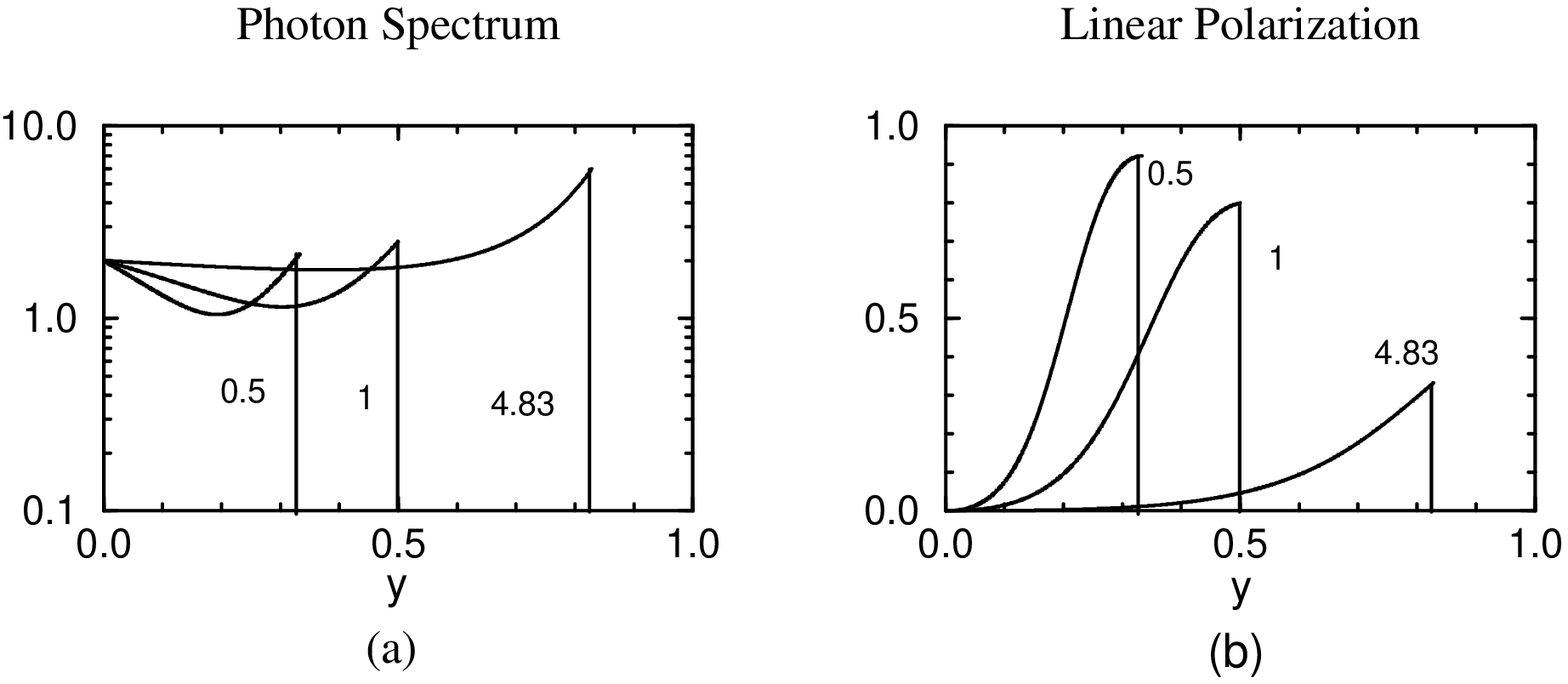,width=15cm,height=11cm}\hss}
\end{figure}

\vskip 3cm
\begin{center}
{\bf\large Figure~8}
\end{center}

\newpage
\mbox{ }
\vskip 3cm
\begin{figure}[h]
\hbox to\textwidth{\hss\epsfig{file=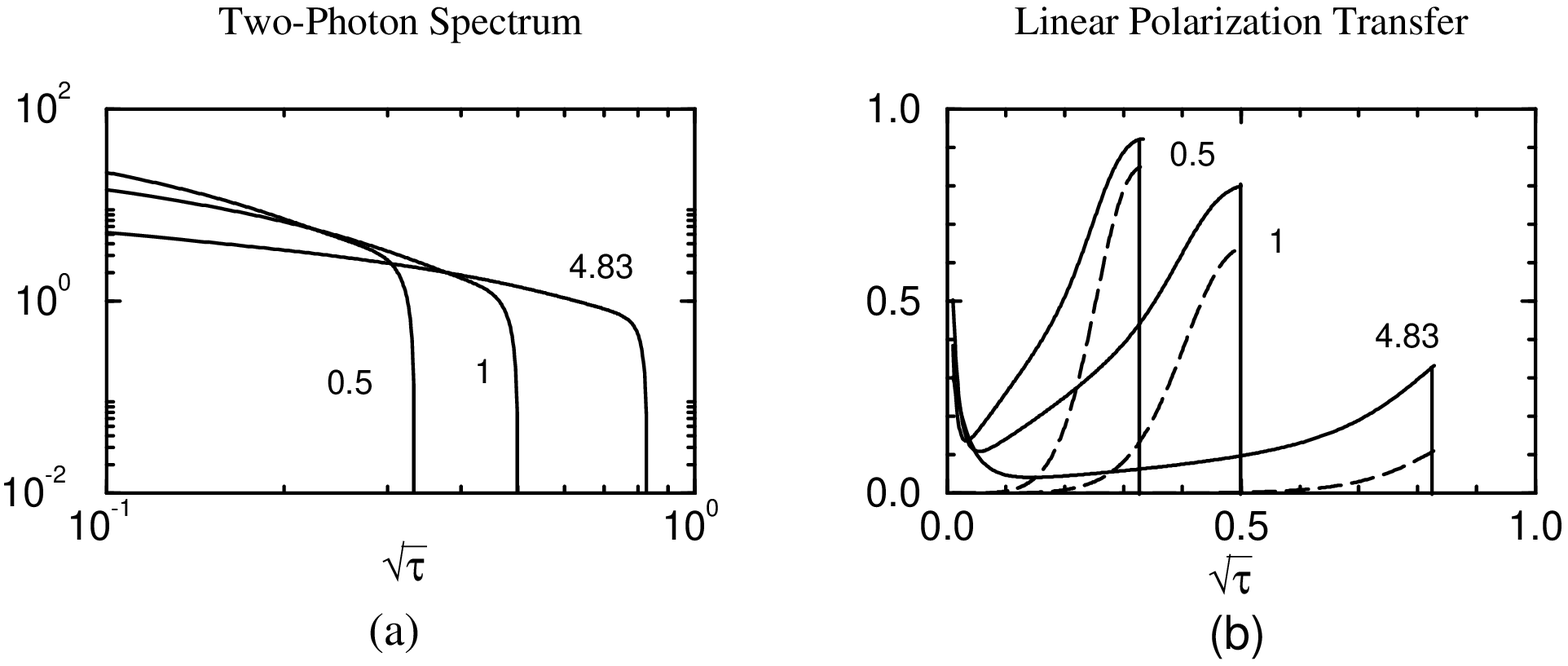,width=15cm,height=11cm}\hss}
\end{figure}

\vskip 3cm
\begin{center}
{\bf\large Figure~9}
\end{center}

\newpage
\mbox{ }
\vskip 3cm
\begin{figure}[h]
\hbox to\textwidth{\hss\epsfig{file=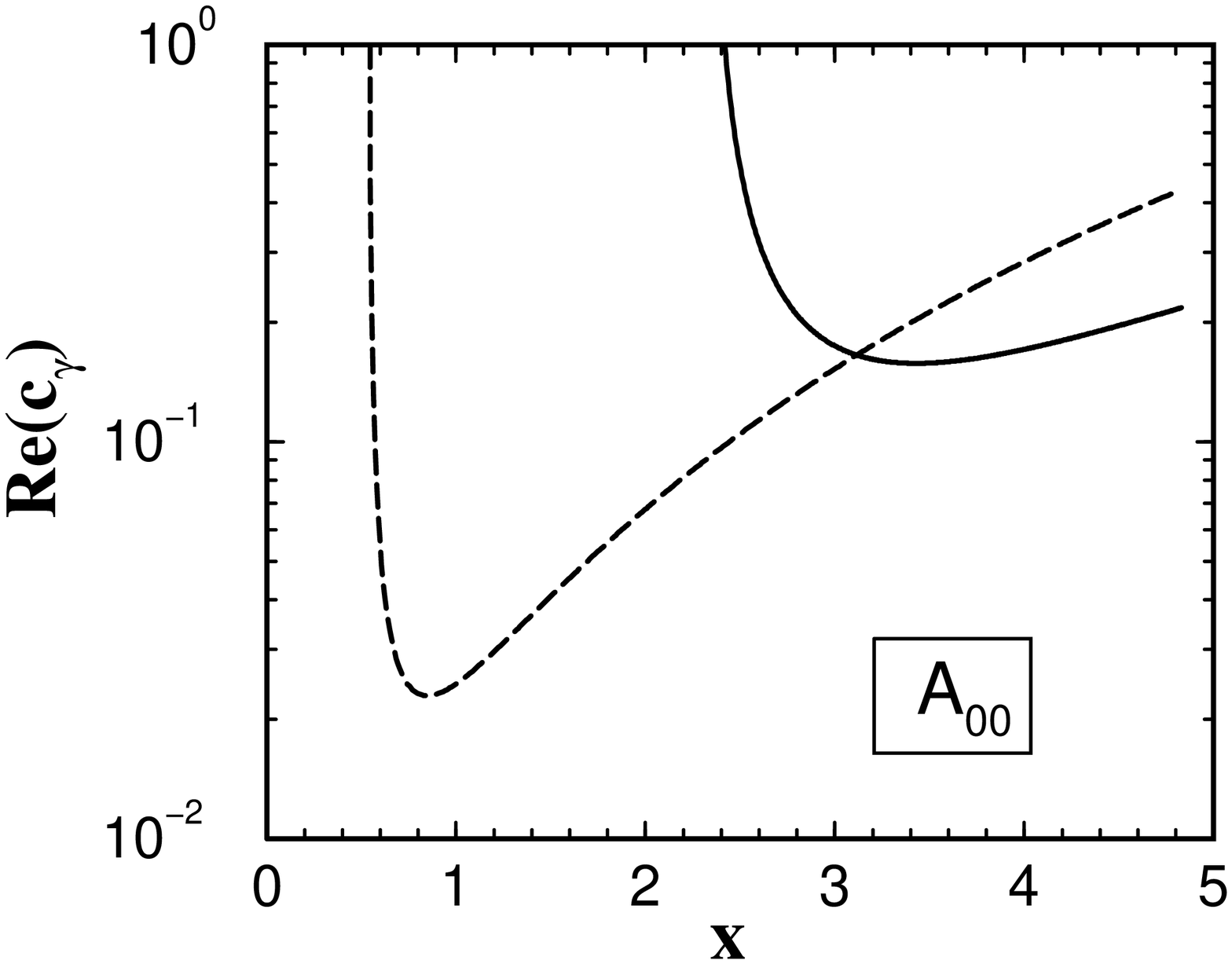,width=15cm,height=11cm}\hss}
\end{figure}

\vskip 3cm
\begin{center}
{\bf\large Figure~10}
\end{center}

\end{document}